\documentclass[twocolumn,superscriptaddress,longbibliography,aps,prl,preprintnumbers,floatfix]{revtex4-2}

\usepackage{dcolumn}
\usepackage{graphicx}
\usepackage{amsmath}
\usepackage{amssymb}
\usepackage{color}
\usepackage{bm} 
\usepackage{flafter}
\usepackage{epsfig}
\usepackage{epstopdf}
\usepackage{float}
\usepackage{multirow}
\usepackage[usenames,dvipsnames]{xcolor}
\usepackage[normalem]{ulem}

\usepackage[urlcolor=blue,colorlinks=true,citecolor=blue,linkcolor=blue,pdfstartview={FitH},bookmarks=false]{hyperref}

\definecolor{jlblue}{rgb}{0.2, 0.5, 0.7}
\definecolor{pjgreen}{rgb}{0.2,0.7,0.2}
\newcommand{\RKK}[1]{\textcolor{black}{#1}}
\newcommand{\RK}[1]{\textcolor{black}{#1}}

\begin{document}

\title{Thermal and Magnetoelastic Properties of the  van der Waals Ferromagnet Fe$_{3-\delta}$GeTe$_2$: Anisotropic  Spontaneous Magnetostriction and Ferromagnetic Magnon Excitations}

\author{Reinhard K. Kremer}
\email{R.Kremer@fkf.mpg.de}
\affiliation{Max Planck Institute for Solid State Research,
             Heisenbergstrasse 1, D-70569 Stuttgart, Germany}

\author{Eva Br\"ucher}
\email{E.Bruecher@fkf.mpg.de}
\affiliation{Max Planck Institute for Solid State Research,
             Heisenbergstrasse 1, D-70569 Stuttgart, Germany}

\date{\today}

\begin{abstract}
By determining the lattice parameters as a function of temperature of the hexagonal van der Waals ferromagnet Fe$_{2.92(1)}$Ge$_{1.02(3)}$Te$_2$  we obtain the temperature dependence  \RKK{of the spontaneous in-plane} magnetostriction in the ferromagnetic and the linear thermal expansion coefficients in the paramagnetic state. The spontaneous magnetostriction is clearly seen in the temperature dependence of the in-plane lattice parameter $a(T)$, but less well pronounced perpendicular to the planes along $c$.
Below $T_{\rm C}$ the spontaneous magnetostriction follows the square of  the magnetization and leads to an expansion of the hexagonal layers.  Extrapolating to $T\rightarrow$ 0~K we obtain a spontaneous \RKK{in-plane saturation magnetostriction of $\lambda_{{\rm sp},a}(T \rightarrow 0) \approx-220 ~\times~10^{-6}$.}
In the paramagnetic state the linear thermal expansion coefficients amount to 13.9(1)$\times$10$^{-6}$~K$^{-1}$ and to 23.2(2)$\times$10$^{-6}$~K$^{-1}$ for the in-plane and out-of-plane direction, respectively, indicating  a linear volume thermal expansion coefficient of 50.8(4)$\times$10$^{-6}$K$^{-1}$ which we use to estimate the volume thermal expansion contribution to the heat capacity determined at constant pressure.
A Sommerfeld-type linear term in the low-temperature heat capacities can be quantitatively ascribed to 2dim ferromagnetic magnon excitations.
\end{abstract}


\maketitle

\section{Introduction}
Two-dimensional (2dim) van der Waals (vdW) ferromagnets with sufficiently high Curie temperatures lately have attracted special attention with respect to their applicability in modern nanoelectronic and spintronic devices.
Among a number of prominent systems, the  ternary iron germanium telluride, Fe$_{3-\delta}$GeTe$_2$ ($\delta \approx 0.1$) (FGT), with
a Curie temperature, $T_{\rm C}$, close to 220~K has been intensively investigated.
After the  initial synthesis of polycrystalline  specimen and the basic  structural and magnetic characterization,\cite{Deiseroth2006}
accessibility of larger crystalline samples
enabled a broad variety of experiments with a focus on  mono- and multilayer samples. These proved FGT  to be a highly promising platform for studying the complex interplay of magnetic and electronic properties in reduced dimensions up to the point of monolayer devices.

FGT exhibits a number of intriguing magneto-electronic properties: Below $T_{\rm C}$ FGT is an itinerant ferromagnet.\cite{Deiseroth2006,BinChen2013} Though, an antiferromagnetic transition at $\sim$150~K has also been conjectured.\cite{Yi2017} Using density functional calculations Jang \textit{et al.} concluded that Fe defects and hole doping are the key to drive FGT into the ferromagnetic phase, whereas they proposed that  stoichiometric FGT to be antiferromagnetic.\cite{Jang2020}

A large anomalous Hall effect already at small polarizing external magnetic fields has been found for  bulk samples.\cite{Tan2018,Wang2018,Kim2018,Liu2018}
Exfoliated monolayers still show uniaxial anisotropy and robust ferromagnetism, however with $T_{\rm C}$ reduced to $\sim$130~K.\cite{Zhuang2016,Fei2018}
Broadest interest attracted the  electrolyte gating experiments by Deng \textit{et al.}. They  showed that already small gate voltages applied to flakes of FGT with Li$^{+}$ intercalated in between the Te double layers can tune $T_{\rm C}$  from 100~K up to room temperature.\cite{Deng2018} Laser-induced spin and charge photocurrents in single-layer FGT were predicted by first-principles calculations suggesting applications of FGT in opto-spintronics.\cite{Merte2021}
N\'{e}el- or Bloch-type skyrmions have been reported to occur in FGT monolayers.\cite{Ding2020,Meijer2020,Wu2020,Yang2020,Park2021}  Xu \textit{et al.} argued that unusual terms in the Hamiltonian connecting four spins can account for these observations.\cite{Xu2022}
Recently, successful generation and manipulation of terahertz spin-current has been reported by Chen \textit{et al.}.\cite{Chen2022}

Despite the booming interest in the magnetic and magnetoelectric properties of FGT, the lattice properties and especially magnetoelastic coupling still remained  scarcely explored.
By using density-functional theory Zhuang \textit{et al.} found that the orbital moment of the Fe atoms is sizeable, causing  a large magnetic anisotropy energy increasing with tensile strain, and  a large in-plane magnetostrictive  coefficient of -559$\times$10$^{-6}$ for monolayer FGT.\cite{Zhuang2016}
In Raman spectroscopy measurements Milosavljevi\'{c} \textit{et al.} observed fingerprints of spin-phonon coupling at around 152~K and also anomalous behavior of the Raman frequencies and mode linewidths at the ferromagnetic transition.\cite{Popovic2019}

FGT crystallizes in the hexagonal crystal system  with lattice parameters $a\approx$ 3.99~\AA\ and $c\approx$ 16.34~\AA.
The crystal structure of FGT contains slabs of Fe$_3$Ge   sandwiched by vdW bonded Te double layers.\cite{Deiseroth2006} The Fe atoms occupy two different crystallographic sites.  Notably, the Fe2  site (Wyckoff position 2c) in samples prepared under stoichiometric conditions is not fully occupied (typically $\delta \sim$ 0.1), whereas the Fe1 site, within error bars, \RKK{exhibits no deficiency}.  Increasing $\delta$, i.e. enlarging the Fe2 deficiency reduces $T_{\rm C}$. The in-plane lattice parameter $a$ decreases with increasing $\delta$, whereas the out-of-plane lattice parameter $c$ increases for larger Fe2 deficiency.\cite{May2016} Strong uniaxial anisotropy aligns the Fe magnetic moments along the $c$ easy axis making FGT an auspicious platform for magnetic data storage devices.\cite{Verchenko2015,Zhu2016,Zhuang2016,Brito2016} The ordered magnetic moments  of the Fe atoms for $\delta$ = 0.1 amount to 2.2~$\mu_{\rm B}$ for the Fe1 and 1.5~$\mu_{\rm B}$ for the Fe2 crystallographic sites, respectively.\cite{May2016} For
$\delta \sim$ 0.25 both ordered moments converge to a common value of $\sim$~1.4~$\mu_{\rm B}$.\cite{May2016,Zhuang2016}

Here, we report on measurements of the zero-field spontaneous magnetostriction  when FGT enters into the ferromagnetic state. Magnetostriction, i.e. the deformation of the geometrical shape  of a ferromagnetic specimen during the magnetization process can be evoked e.g. by an external magnetic field. Especially magnetic field induced magnetostriction is of great technological importance but also  fundamental to model  the magnetization process itself and the formation of the domain structure.
Magnetostriction induced by an external magnetic field depends on the orientation of the external magnetic field with respect to the orientation of domain magnetization and the direction of the exchange interaction between the magnetic moments.\cite{Chikazumi1997}
Spontaneous magnetostriction where the shape change, $\lambda$=$\delta l(T)/l(T)$, is initiated by the increase of the spontaneous magnetization $M(T<T_{\rm c},H=0)$ below  $T_{\rm C}$ in vanishing external magnetic field $H$ i.e. by the internal ferromagnetic domain formation can provide very valuable information, e.g. about spontaneous reorientation processes and changes in the domain structure. Technologically magnetostrictive effects may become important for the interaction of FGT monolayers deposited on substrates.

In addition, we review preceding heat capacity experiments which have found large   linear terms in the low and high temperature specific heats which the authors attributed to magnetic contributions.\cite{BinChen2013} Especially  in the paramagnetic regime, the published heat capacity data substantially overshoot the Dulong-Petit limit, suggesting a critical reconsideration of the thermal properties. By using the linear thermal volume expansion coefficients in the paramagnetic state to determine the  lattice expansion contribution to the heat capacities measured at constant pressure, $C_p$, we revise these findings.

The Sommerfeld-like linear term in the low-temperature heat capacities can be quantitatively attributed to 2dim ferromagnetic magnon excitations. A comparison with the spin stiffness constants found by inelastic neutron scattering gives quantitative agreement.

\section{Experimental}
Samples of FGT were prepared from stoichiometric mixtures of powders of the elements, Fe   (Alfa Aesar, purity 99.998\%), Ge  (Thermo Fisher, purity 99.999\%) and Te  (Thermo Fisher, purity 99.999\%) using a minute amount of iodine as mineralizer. The starting materials  were thoroughly mixed  and reacted in evacuated quartz glass tubes in a two-zone furnace  heated to temperatures between 750$^{\circ}$C and  650$^{\circ}$C. Phase purity and composition of the samples was checked by energy-dispersive spectroscopy (EDX) employing a Tescan Vega4 LMU  SEM equipped with an Oxford X-Max$^N$N20 detector and by x-ray  powder diffraction (XRPD) using  Mo$K \alpha_1$ and Cu$K \alpha_1$ radiation.
The magnetic properties of the sample were determined  by powder and single crystal dc magnetization (MPMS, Quantum Design)  and  specific heat measurements (PPMS, Quantum Design).

The lattice parameters of a polycrystalline sample of FGT were determined \RKK{in zero external magnetic field} as a function of temperature from Rietveld profile analysis\cite{Fullprof} of XRPD patterns collected on a  Bruker D8 Discovery x-ray diffractometer  (Bragg-Brentano scattering geometry) using Cu$K \alpha_1$ radiation. A PHENIX (Oxford Cryosystems) closed cycle  cooling cryostat  was used to set the temperature of the sample. Each XRPD pattern was collected at stabilized temperature.
The particle size of the powder was adjusted to  63~$\mu$m or less by  straining the powder through sieves of the respective mesh size. The  powder sample was thermally anchored with ApiezonN vacuum grease to the sample holder platform, equipped with an inlay of a  Si wafer specially oriented to suppress background scattering from the sample holder.

\section{Results and Discussion}
\subsection{Sample Characterization}
\begin{figure}
	\centering
    \includegraphics[width=6.5cm]{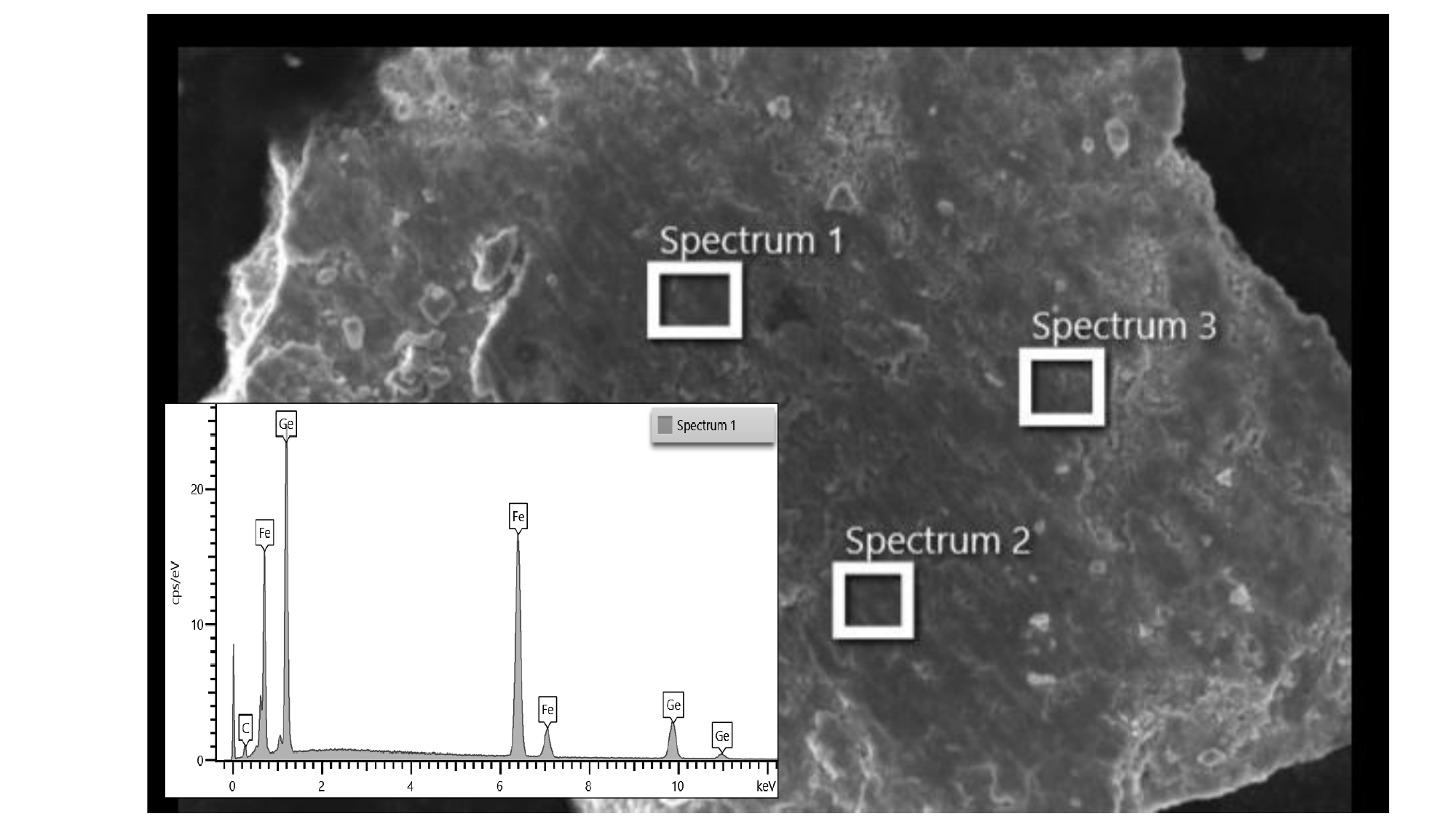}
\caption{SEM image of a crystal of our sample with the rectangles marking the areas where EDX element analysis was performed. The variation of element concentration across the analyzed areas amounted to less than 0.5\%, 1.4\% and 0.07\% for Fe, Ge and Te, respectively. \RKK{The lower inset shows the SEM spectrum collected at spot no. 1.}}
	\label{Fig-SEM}
\end{figure}

Figure \ref{Fig-SEM} displays a SEM picture of a typical crystal analyzed with energy dispersive x-ray spectroscopy. Three spots on an $a$-$b$ surface were tested and very good agreement was found for the element concentration, proving homogeneity of the element concentration.
EDX analyses carried out  on  crystals of several other FGT sample preparations indicated  a spread of the  Fe2 concentration of  0.82~$\leq (1-\delta) \leq$~0.92.

\begin{figure}
	\centering
    \includegraphics[width=8.5cm]{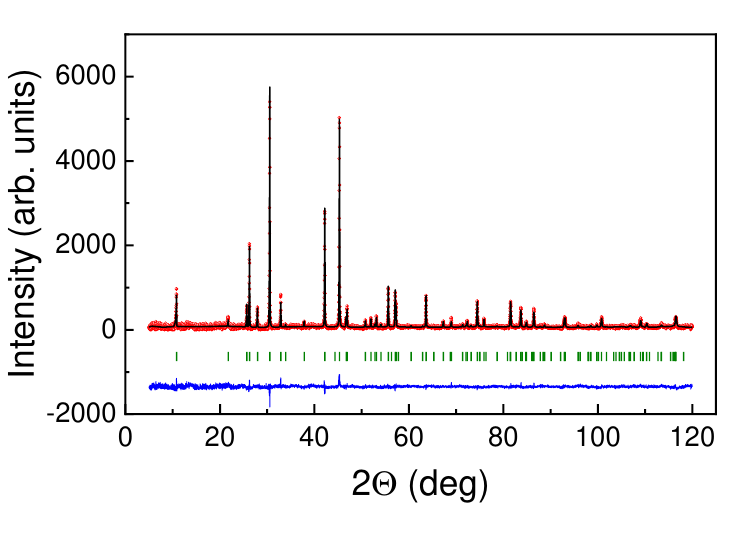}
\caption{(color online) XRD pattern of FGT (Cu$K\alpha_1$ radiation) collected at $295$~K. The red circles represent the experimental data, the solid black line show the result of the Rietveld profile refinement and the vertical green bars mark the positions of the Bragg reflections used to calculate the refined pattern.}
	\label{Fig-1}
\end{figure}

XRPD  on crushed crystals and polycrystalline samples of FGT   proved phase purity of our specimen. Figure \ref{Fig-1} displays a typical room-temperature  XRPD pattern analyzed  with a Rietveld profile refinement assuming the structure model proposed by Deiseroth \textit{et al.} (space group $P$6$_3$/$mmc$, no.~194).\cite{Deiseroth2006} At room temperature the refined lattice parameters amount to $a$ = 3.99939(8) \AA\ and $c$ = 16.3225(4) \AA, in good agreement with the values reported by Deiseroth \textit{et al.}.
In the refinements the Ge and Fe1 lattice sites were assumed to be fully occupied, whereas for Fe2 the occupancy converged to 0.918(11),in good accord with the EDX result. At room temperature the $z$-positional parameter of the Te and Fe1 atoms (Wyckoff positions 4f and 4e, respectively) were refined to 0.5902(2) and 0.6702(3), also in good agreement with the values obtained from single crystal structure determination.\cite{Deiseroth2006}

\begin{figure}[H]
	\centering
    \includegraphics[width=8.5cm]{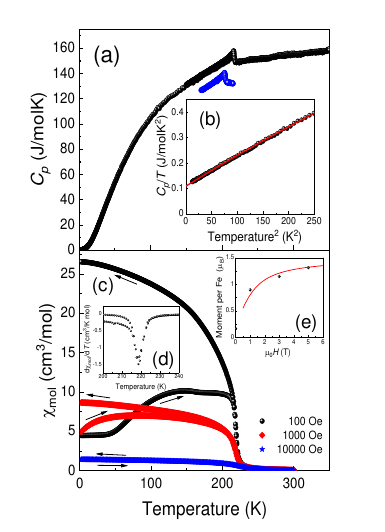}
\caption{(color online) \RKK{(upper panel, black circles) (a) Molar heat capacity of a crystal of Fe$_{2.92(1)}$Ge$_{1.02(3)}$Te$_2$. The blue dots exemplify the reduction of the Curie temperature for a sample with element composition Fe$_{2.85}$GeTe$_2$  (data  downshifted by 10 J/molK for better comparison). (b) displays the power law behavior at low temperatures.(lower panel) (c) zfc and fc magnetic susceptibilities (indicated by the arrows) of the FGT sample of composition Fe$_{2.92(1)}$Ge$_{1.02(3)}$Te$_2$ measured with external magnetic fields of 100~Oe, 1~kOe and 10~kOe. The inset (d) displays the derivative of the magnetization measured with an external field of 100~Oe as a function of temperature. The (upper) inset (e) shows the magnetic moment per Fe atom. The (red) solid line is a guide to the eye.}
}
	\label{Fig-2}
\end{figure}

The Curie temperatures of the FGT samples were determined from heat capacity and magnetization measurements. Figure \ref{Fig-2} summarizes the magnetic susceptibilities and the specific heat of the sample with composition Fe$_{2.92(1)}$Ge$_{1.00(3)}$Te$_2$.
The magnetization data exhibit the characteristic splitting of the zero-field cooled (zfc) and field - cooled (fc) branches \RKK{reported} before.\cite{Deiseroth2006} The zfc - fc hysteresis closes with increasing magnetic field.
The steepest descent of the magnetization is found at $\sim$218~K (see inset (b) in Figure \ref{Fig-2}) where also the $\lambda$-type anomaly in the specific heat is observed (Figure \ref{Fig-2}(d)).
At room temperature the heat capacity linearly approaches a value of $\sim$156~J/molK, consistent with the Dulong-Petit value of 5.92$\times$3$R$, where $R$ is the molar gas constant, but substantially lower than the findings reported by Bin Chen \textit{et al.}.\cite{BinChen2013} The linear increase of the heat capacity will be discussed in detail below.
At low temperatures in the ferromagnetic state the heat capacities follow a power law (Figure \ref{Fig-2}(e))
\begin{equation}
C_{{mol}}/T = \gamma + \beta T^2,
\label{Sommerfeld}
\end{equation}
with $\gamma$ = 109(1)~mJ/molK$^2$, close to what has been observed by Bin Chen \textit{et al.}.\cite{BinChen2013} The slope $\beta$ = 1.25(6) mJ/molK$^4$ implies a Debye temperature, $\Theta_{\rm Deb}(T \rightarrow$ 0~ K) of 210(2)~K. The \RKK{majority of the} linear contribution to the heat capacity \RKK{can be} ascribed to two-dimensional ferromagnetic magnons, as will be analyzed in detail below.

\begin{figure}[H]
	\centering
    \includegraphics[width=8.0cm]{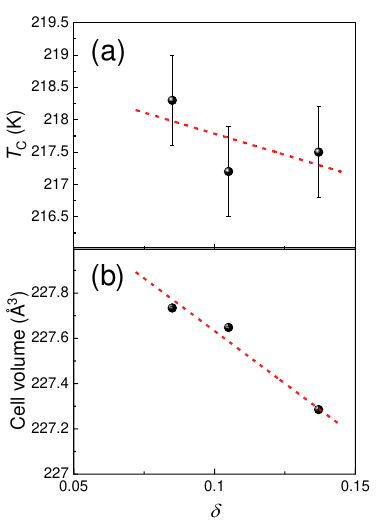}
\caption{(color online) (a) Curie temperature, $T_{\rm C}$ as a function of the Fe2 deficiency, $\delta$ of our samples. $\delta$ was determined from the Rietveld refinements of the XRPD patterns collected using Mo$K\alpha_1$ radiation and from EDX analyses.  (b) Volume of the crystallographic unit cell as a function of $\delta$.  \RKK{The (red) dashed  lines are guides to the eye.}}
	\label{TcvsConc}
\end{figure}

The Curie temperature $T_{\rm C}$ depends on the Fe2 content. An increase of $\delta$  leads to a reduction of the Curie temperature (see blue dots in  Figure \ref{Fig-2}(a)). Figure \ref{TcvsConc}(a) displays the variation of the Curie temperature as a function of the composition around $\delta \approx$~0.1.  A decrease of $T_{\rm C}$ is paralleled by a decrease  of the volume of the crystallographic unit cell (Fig. \ref{TcvsConc}(b)), the latter induced by the decrease of the in-plane lattice parameter $a$.

\subsection{Spontaneous Magnetostriction}
\begin{figure}[H]
	\centering
    \includegraphics[width=9.5cm]{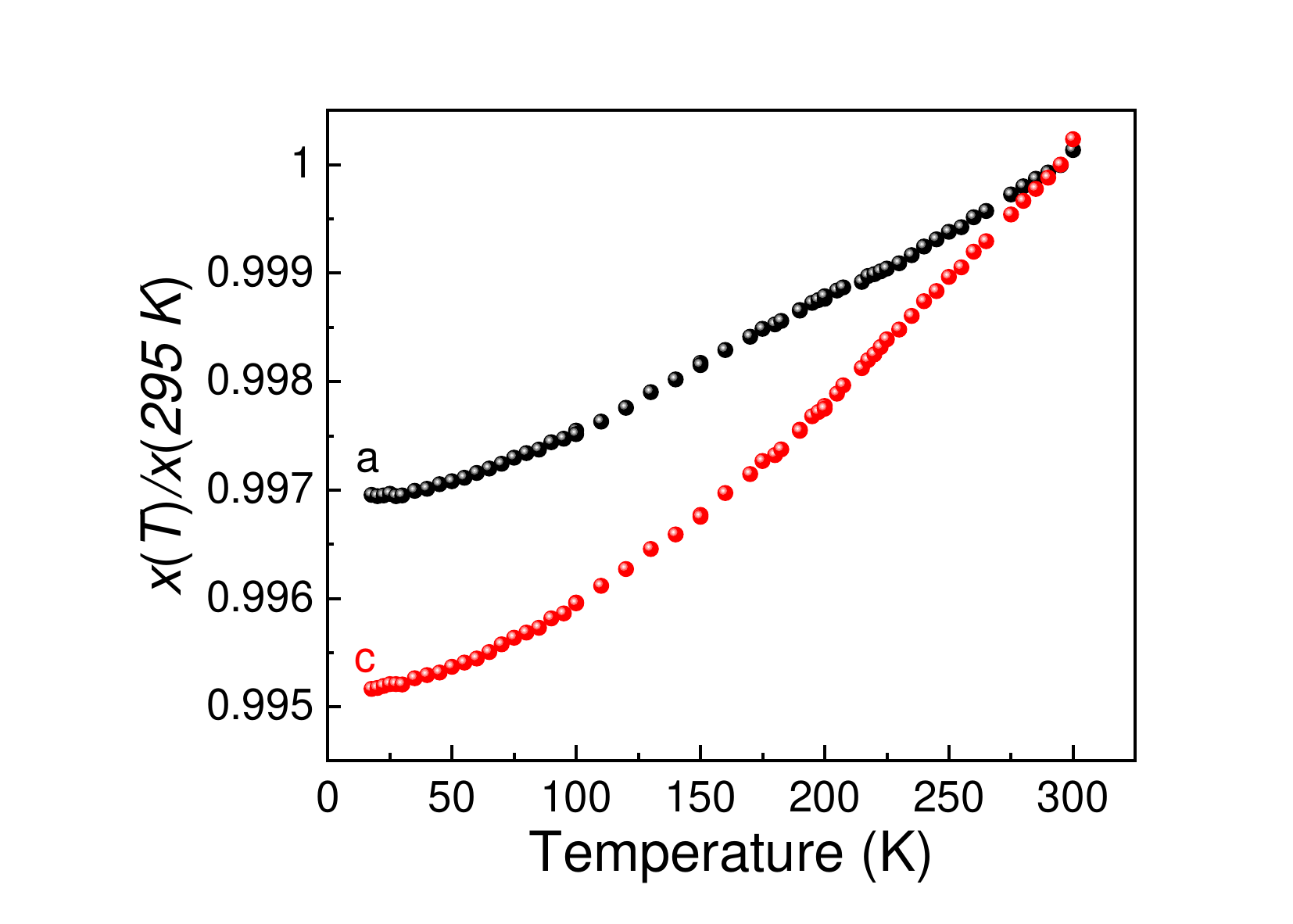}
\caption{(color online) Temperature dependence of the lattice parameters $a^*(T)$=$a(T)$/$a(295K)$  and  of $c^*(T)$=$c(T)$/$(295K)$.}
	\label{Fig-Rel}
\end{figure}

\begin{figure}
	\centering
    \includegraphics[width=8.5cm]{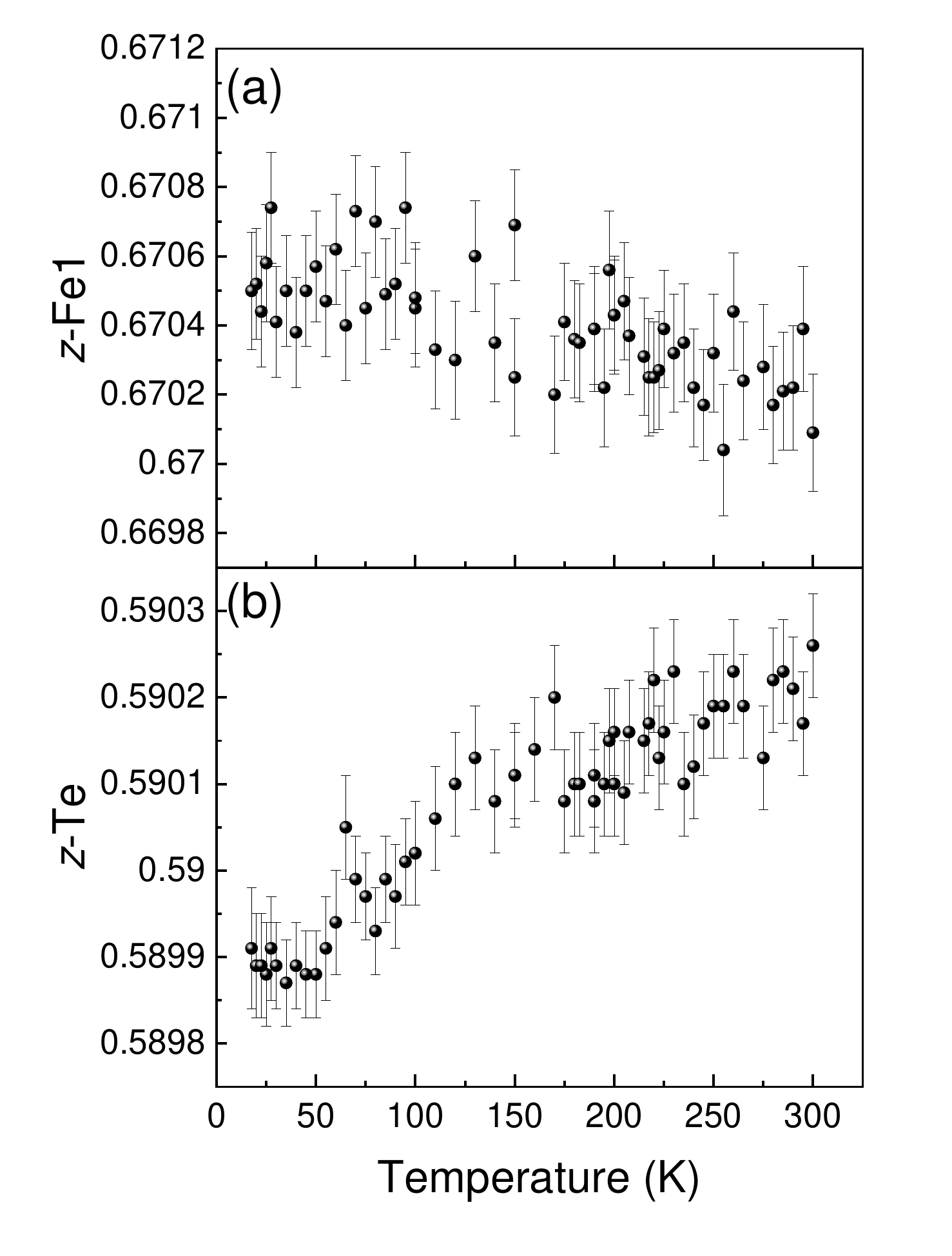}
\caption{$z$ positional parameters of the (a) Fe1 and (b) the Te atoms (Wyckoff positions 4e and 4f, respectively) of Fe$_{2.92(1)}$Ge$_{1.02(3)}$Te$_2$ as a function of temperature.}
	\label{zParam}
\end{figure}

\RKK{A fraction of the same sample with composition Fe$_{2.92(1)}$Ge$_{1.02(3)}$Te$_2$ which had never been exposed to an external field was subsequently used for the temperature dependent XRPD measurements.}
Figure~\ref{Fig-Rel} displays the  \RKK{reduced} lattice parameters $a^*(T)$=$a(T)$/$a$(295~K) and  $c^*(T)$=$c(T)$/$c$(295~K), relative to their room temperature values, as derived from the Rietveld profile refinements of the temperature dependent XRPD patterns. With decreasing temperature the relative contraction of the lattice parameters  perpendicular to the layers is about a factor of two larger than the in-plane contraction.
In the paramagnetic regime  the cell volume and the lattice parameters decrease linearly with temperature with  rates
($dV_{\rm cell}/dT$)/$V_{\rm cell}$(295~K) = -50.8(4)$\times$10$^{-6}$~K$^{-1}$, and -13.9(2)$\times$10$^{-6}$~K$^{-1}$, and  -~23.2(2)$\times$10$^{-6}$~K$^{-1}$, for $a$ and $c$, respectively.
Figure \ref{zParam} displays the temperature dependence of the $z$-positional parameters of the Te and the Fe1 atoms. Whereas $z$(Fe1) shows a faint increase with decreasing temperature, still within error bars, the $z$ positional parameter of the Te atoms decreases and levels off at low temperatures.

\begin{figure}
	\centering
    \includegraphics[width=9.5cm]{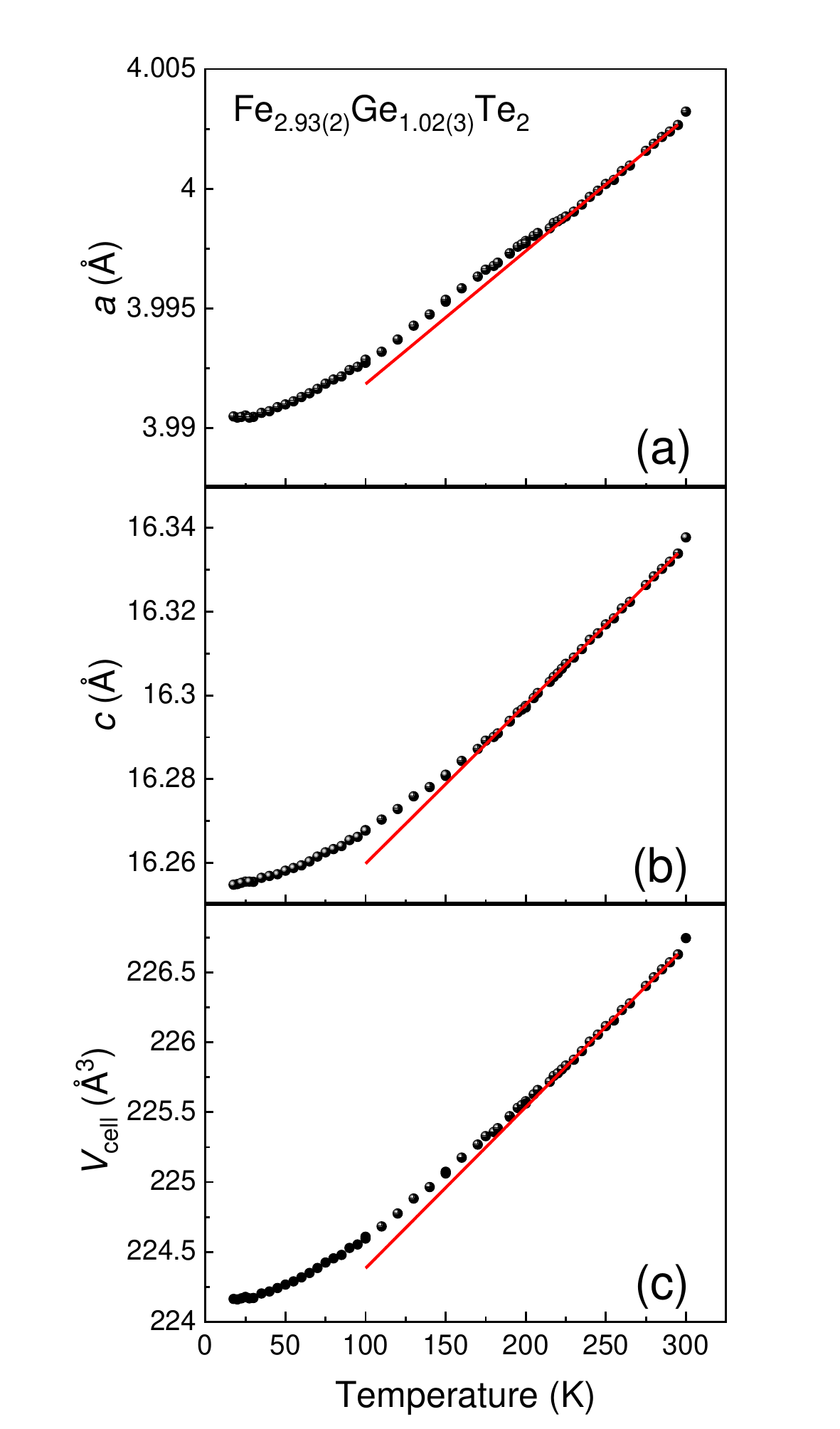}
\caption{(color online) Temperature dependence of the lattice parameters $a$ (a), $c$ (b), and the cell volume (c)  of Fe$_{2.92(1)}$Ge$_{1.02(3)}$Te$_2$. Error bars are of the size of the symbols.
The solid (red) lines represent linear fits between 235~K and 295~K of the temperature dependence of the respective quantities. The relative slopes are given in the text.}
	\label{Fig-3}
\end{figure}

Figure \ref{Fig-3} compiles the volume of the hexagonal unit cell  and the lattice parameters $a(T)$ and $c(T)$ as a function of the temperature.
Whereas the in-plane lattice parameter $a(T)$ exhibits a noticeable shoulder near  $T\sim T_{\rm C}$,  $c(T)$  bents away from the high-temperature linear behavior before leveling off at low temperatures. A noticeable response to the onset of ferromagnetic order is not seen for the out-of-plane lattice parameter $c$.
The shoulder in  $a(T)$ is ascribed to a negative spontaneous in-plane magnetostriction, i.e. an expansion  of the hexagonal planes, induced by ferromagnetic ordering.

\begin{figure}
	\centering
    \includegraphics[width=9.0cm]{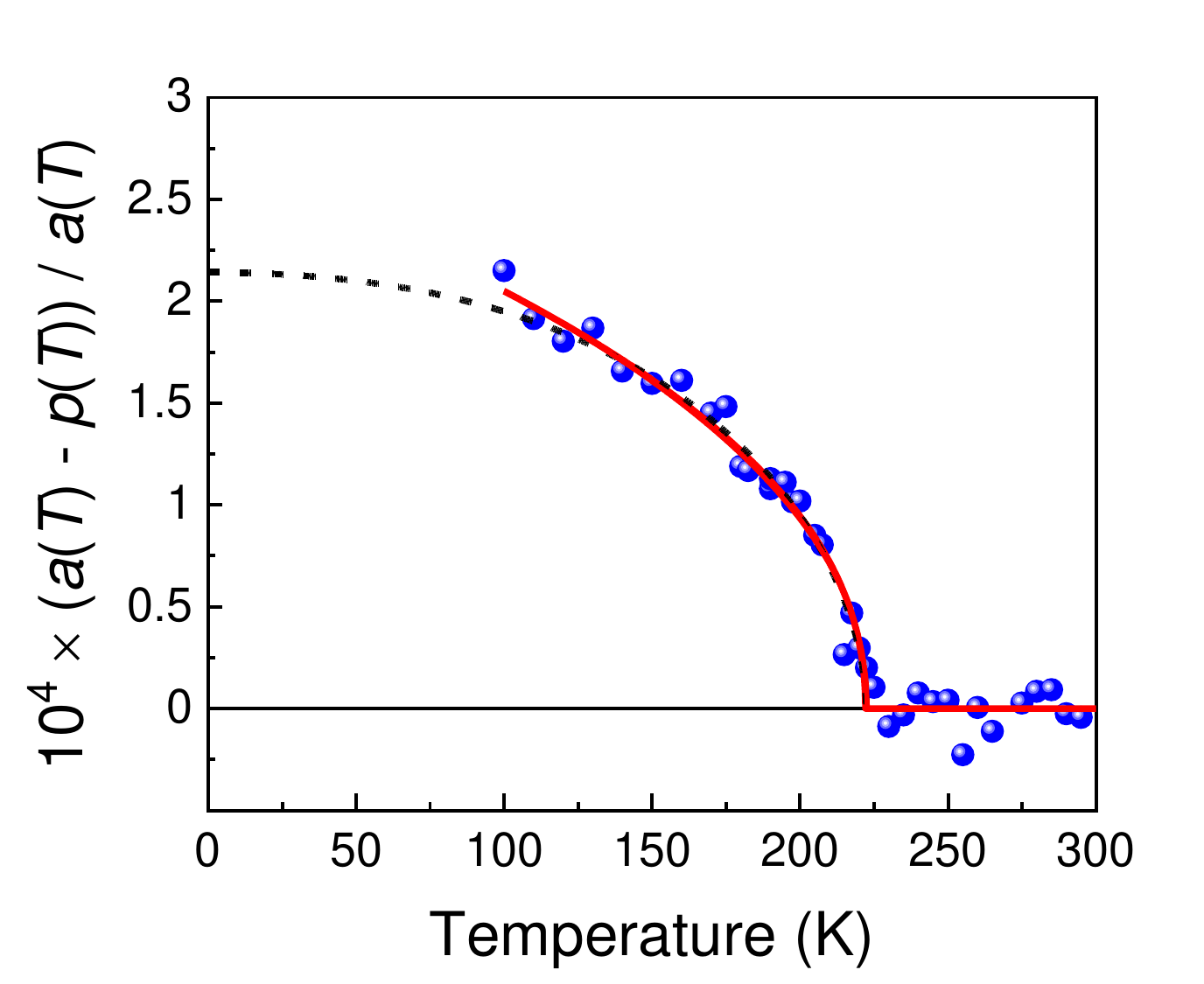}
\caption{(color online) \RK{Spontaneous in-plane magnetostriction of Fe$_{2.92(1)}$Ge$_{1.02(3)}$Te$_2$ as a function of temperature. The red
solid line is a fit to a critical power law, $\tau^{\beta}$ with the reduced temperature  $\tau$ = 1~-~($T/T_{\rm C}$) with a critical temperature, $T_{\rm C}$ 222.4(8) K and a critical exponent $\beta$ = 0.48(2). The black dashed line represents an extrapolation to $T \rightarrow$ 0 K.}}
	\label{Fig-4}
\end{figure}

Extrapolating the linear behavior of $a(T)$ from the paramagnetic regime and subtracting from the data below $T_{\rm C}$, reveals a spontaneous in-plane magnetostriction starting below $\sim$223~K,  which  saturates below $\sim$50~K. The temperature dependence of the spontaneous magnetostriction (see Figure \ref{Fig-4}) is reminiscent of a continuous phase transition with a critical temperature of 223(1)~K, matching  $T_{\rm C}$, as determined from the magnetization and specific heat experiments.
\RK{A tentative extrapolation of the spontaneous magnetostriction to $T \rightarrow$ 0 K  (see the  black dashed line in Fig. \ref{Fig-4})} indicates a spontaneous  in-plane saturation magnetostriction,  $\lambda_{{\rm sp},a}(0))$, of
\begin{equation*}
\lambda_{{\rm sp},a}(T \rightarrow 0) \approx -2.2 \times 10^{-4}.
\label{Kuzmineqn2}
\end{equation*}

Attempts to model the temperature dependence of the lattice parameters  to a Debye function over the whole temperature range lead to unstable fits and were not meaningful. For the same reason the out-of-plane spontaneous magnetostriction was difficult to ascertain since there is no visible anomaly in the temperature dependence of the $c$-lattice parameter at $T_{\rm C}$.

\begin{figure}
	\centering
    \includegraphics[width=8.5cm]{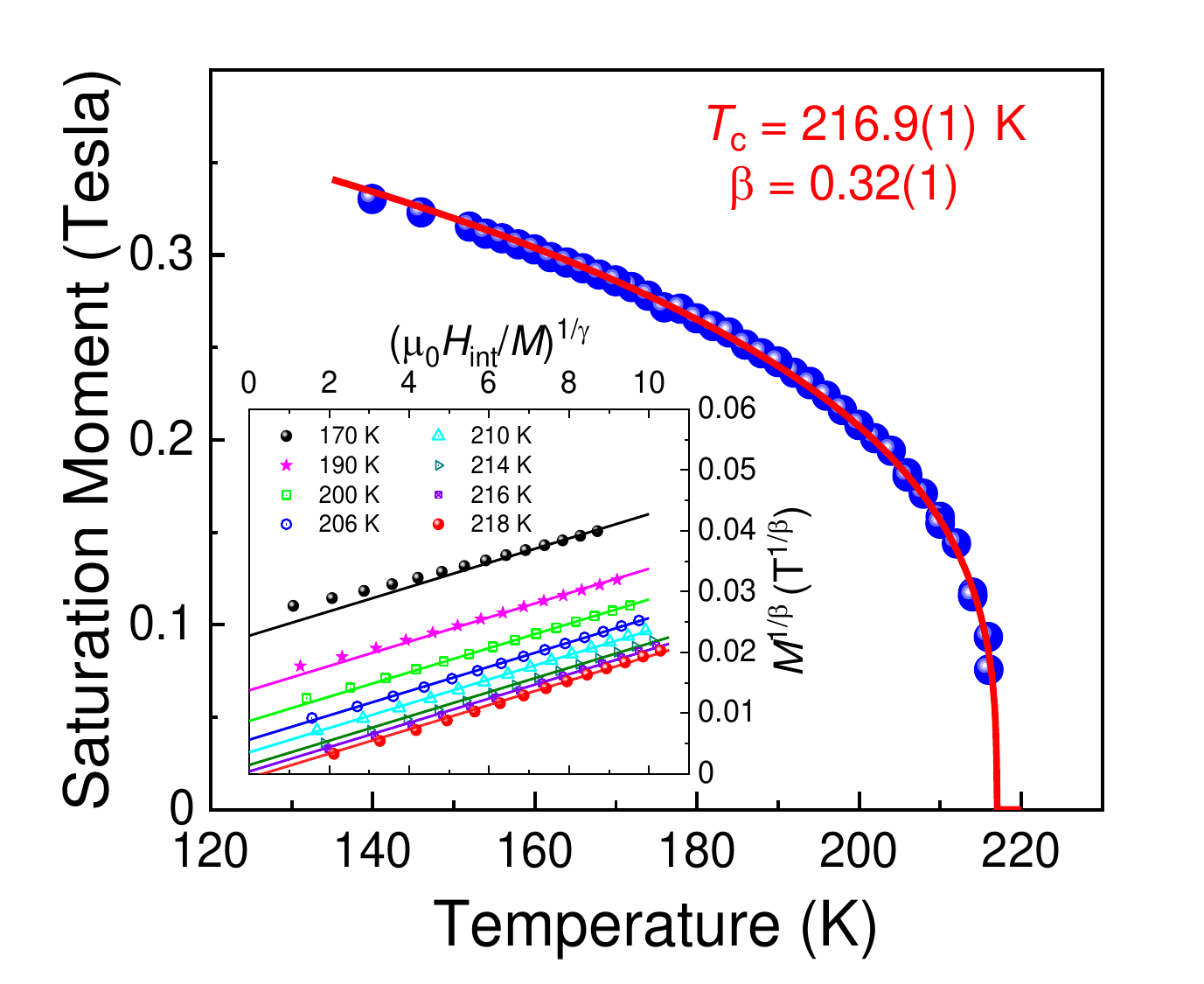}
\caption{(color online) \RK{Temperature dependence of the zero-field spontaneous magnetization of  Fe$_{2.92(1)}$Ge$_{1.02(3)}$Te$_2$ derived from modified Arrott Belov plots displayed (lower inset)
adopting  critical exponents  $\beta$ = 0.33 and $\gamma$ = 1.38. The main frame shows a fit of the magnetization with a power-law temperature dependence (solid red) line according to eq. (\ref{Critical}).}}
	\label{Arrott}
\end{figure}

\RK{In order to relate the spontaneous magnetostriction to the temperature dependence of the magnetization we carried out isothermal magnetization measurements on a thin crystal platelet of Fe$_{2.92(1)}$Ge$_{1.02(3)}$Te$_2$ with the magnetic field aligned along the $c$-axis. After a correction for the demagnetizing field, we derived the temperature dependence of the zero-field magnetization from modified Arrott Belov plots\cite{Reisser1,Reisser2} (see inset in Fig. \ref{Arrott}).}
\RK{The zero-field magnetization can be very well fitted to a critical power law  (see  Fig. \ref{Arrott}) given by}
\begin{equation}
M(T) = M_0 (1 - T/T_{\rm C})^{\beta},
\label{Critical}
\end{equation}
\RK{with a critical exponent $\beta$ = 0.32(1), consistent with findings reported earlier by  Liu et al..\cite{Liu1,Liu2}}

\begin{figure}
	\centering
    \includegraphics[width=8.5cm]{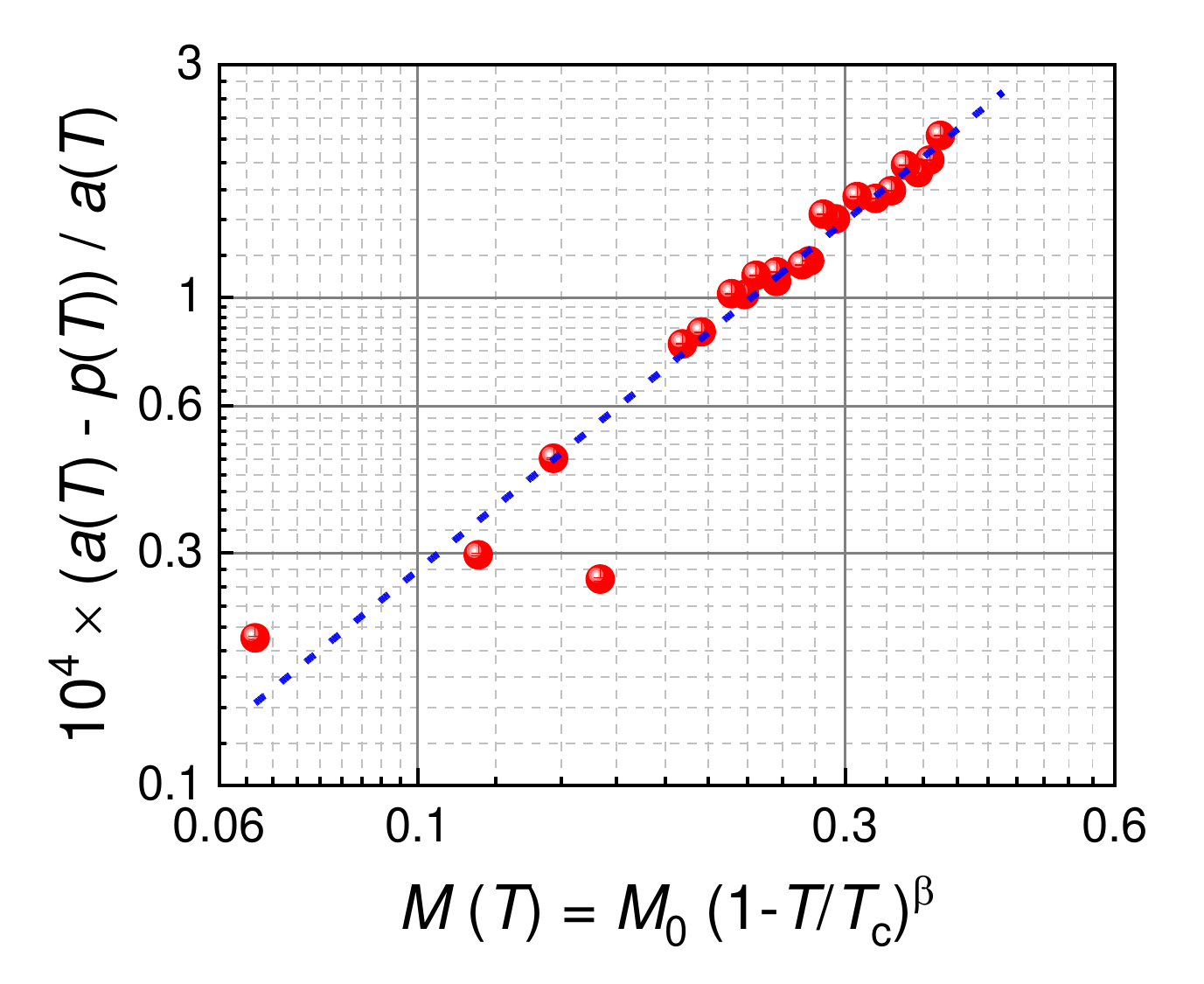}
\caption{(color online) \RK{Log-log plot of the spontaneous in-plane magnetostriction as a function of the magnetization. The blue dashed line corresponds to a power law with an exponent 1.5.}}
	\label{LogLog}
\end{figure}

\RK{Fig. \ref{LogLog} displays the temperature dependence of the spontaneous magnetostriction as a function of the magnetization.}
\RK{The log-log plot reveals a power law behavior of the spontaneous magnetostriction according to}
\begin{equation}
\lambda_{\rm sp,a} (T) =  [M_0 (1 - T/T_{\rm C})^{\beta}]^{1.5(1)}.
\label{LambdavsM}
\end{equation}
\RK{Taking into consideration the critical exponent $\beta$ = 0.32 for the magnetization, eq.(\ref{LambdavsM}) indicates a critical exponent for the spontaneous in-plane magnetostriction of $\sim$0.48, i.e. close to the critical exponent expected for mean field behavior.}

\subsection{Thermal Lattice Expansion Contribution to the Specific Heat}
\RKK{In a first investigation of the heat capacities of FGT Bin Chen \textit{et al.} had found large linear contributions at low and high temperatures, hitherto no conclusively explained.\cite{BinChen2013}}  At room temperature they found a value of $\sim$200 J/molK and  a stark linear increase above $T_{\rm C}$, which they  attributed to electronic and magnetic contributions.\cite{BinChen2013}

In contrast to these earlier findings our  heat capacity data measured at constant pressure, $C_p$,  (see Fig. \ref{Fig-2}(d))  tend to a value of $\sim$156 J/molK, only moderately exceeding  the Dulong-Petit limit for the heat capacity at constant volume, $C_v (T \rightarrow \infty) \times 3NR =$ 148 J/molK, with $N$ = 5.92 being the number of atoms per formula unit and $R$ the molar gas constant. At 350~K our $C_p$ data attain a value of $\sim$158~J/molK,
substantially lower than that reported by Bin Chen \textit{et al.}.
\begin{figure}
	\centering
    \includegraphics[width=10.5cm]{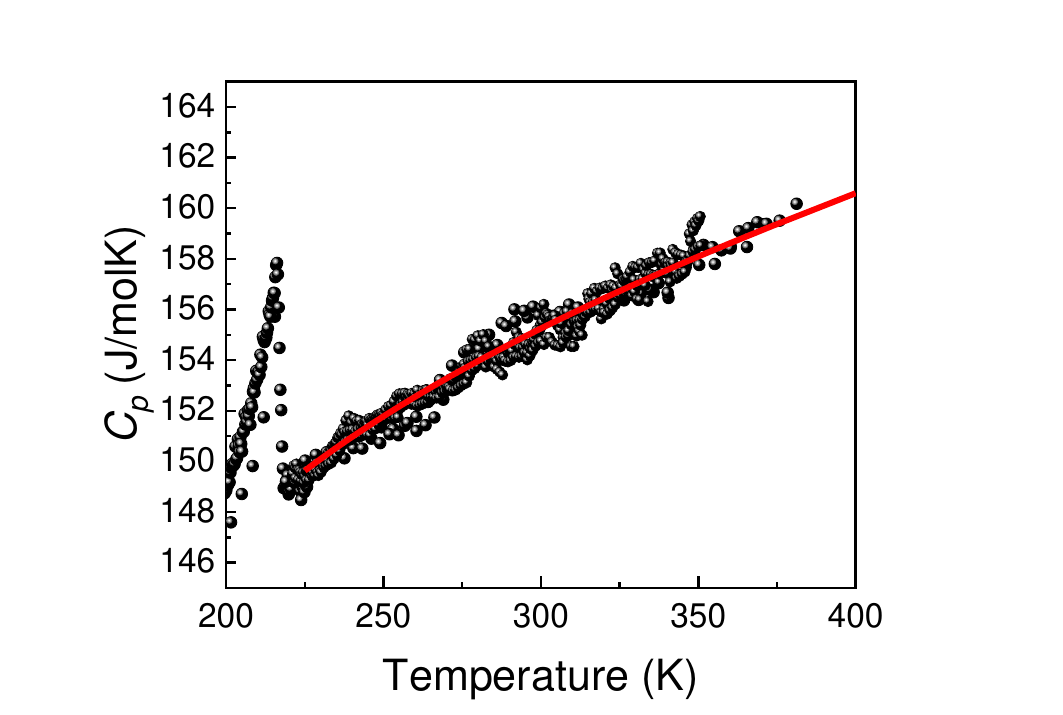}
\caption{(color online) High-temperature heat capacities at constant pressure, $C_p$, of Fe$_{2.92(1)}$Ge$_{1.02(3)}$Te$_2$. \RKK{The solid (red) line is a fit to eq.(\ref{Tolman}) with parameters given in the text.}}
	\label{AllHeat}
\end{figure}

\RKK{In the following, using our thermal expansion data for $T > T_{\rm C}$, we estimate the electronic and the lattice thermal expansion contribution to the heat capacities at high temperatures.}
The difference between the heat capacities measured at constant pressure and at constant volume is given by
\begin{equation}
C_p - C_v =   \alpha_v^2(T)B V_{{mol}} T,
\label{DiffCpCv}
\end{equation}
where $\alpha_v(T)$ is the temperature dependent coefficient of the volume thermal
expansion, $B$ is the bulk modulus and $V_{mol}$ the molar volume.

In the paramagnetic regime above $T_{\rm C}$ and up to room temperature  the lattice expands approximately linearly with increasing temperature. The linear thermal  expansion coefficients amount to $\alpha_a$~=~13.9(1)$\times$10$^{-6}$~K$^{-1}$ and to $\alpha_c$~=~23.2(2)$\times$10$^{-6}$~K$^{-1}$ for the in-plane and out-of-plane direction, respectively,  resulting in a linear volume thermal expansion coefficient of $\alpha_v$~=~50.8(4)$\times$10$^{-6}$.
The in-plane linear thermal expansion coefficient of   Fe$_{2.92(1)}$Ge$_{1.02(3)}$Te$_2$ in the paramagnetic state compares well with that of elementary iron for which an expansion coefficient of 11.6$\times$10$^{-6}$ was observed at room temperature.\cite{Touloukian}

In order to fit  the heat capacity at high temperatures displayed in Figure \ref{AllHeat}  we used a series expansion for the Debye contribution\cite{Tolman1938} and included a  term linear in temperature, $E~ T$,
\begin{equation}
C_p(T) = 5.92\times 3\,R\,\, (1 -
\frac{1}{20}\,\frac{\Theta_\infty^2}{T^2} +
\frac{1}{560}\,\frac{\Theta_\infty^4}{T^4}) + E~T,
 \label{Tolman}
\end{equation}
\noindent with $R$ being the gas constant, and $\Theta_{\infty}$ the Debye temperature.  $E~T$ comprises the  thermal expansion contribution given by Eq. (\ref{DiffCpCv}) but also a linear electronic (Sommerfeld) term $\gamma$. A fit of the heat capacities of several runs (Fig. \ref{AllHeat}) using data for
$T >$ 225 K  yields
\begin{equation*}
E = 37(2)~\rm{mJ/mol K^2},
 \label{HTlinear}
\end{equation*}
and
\begin{equation*}
\Theta_{\infty} = 222(3)~\rm{K},
 \label{Thetainfty}
\end{equation*}
the latter being very close to the Debye temperature $\Theta_{\rm Debye}(T\rightarrow$~0~K) which amounts to 210~K.

The Sommerfeld-term, $\gamma$,  from the heat capacity of the conduction electrons can be estimated from
\begin{equation}
\gamma = \frac{\pi^2 k_{\rm B}^2}{3} N(E_{\rm F}),
\label{Conelec}
\end{equation}
with the Boltzmann constant $k_{\rm B}$ and the electronic density of states $N(E_{\rm F})$ at the Fermi level.
The electronic densities of states for two spin directions at $E_{\rm F}$ per formula unit obtained from density functional calculations  has been reported to 6(2) states/eV.\cite{Kim2021,Dang2023}
Using this value  and  Eq. (\ref{Conelec}) one arrives  at  a conduction electron  term of 14~mJ/molK$^2$ and at a value of 23 mJ/molK$^2$ \RKK{for the thermal expansion contribution}. Adopting the volume thermal expansion coefficient $\alpha_{\rm {Vol}}$ = 50.8$\times$10$^{-6}$~K$^{-1}$ and the molar volume $V_{mol}$ =  6.785$\times$10$^{-5}$ m$^3$mol$^{-1}$ \cite{Deiseroth2006}  results in a Bulk modulus $B$ of
\begin{equation*}
B \sim130(40) \rm{GPa},
 \label{Modulus}
\end{equation*}
wherein the sizeable error bar arises from the uncertainty of the electronic density of states at the Fermi energy.
Our finding for the bulk modulus is by a factor of two larger than the value of $\sim$70~GPa reported in literature\cite{Ohara2020,Dang2023} The difference between our value and the reported results might be partially due to the negligence of possible electron-phonon enhancement of the electronic Sommerfeld term. We also note that for their high-pressure M\"ossbauer measurements O'Hara \textit{et al.} used FGT samples with a $T_{\rm C}$ of 155~K, indicating  a substantial deficit on the Fe2 site which may also affect their value for the bulk modulus.\cite{Ohara2020}

Even accepting the full value of $E \approx$~37~mJ/molK$^2$ as due to the electron contribution, the \RKK{excess} heat capacity of almost 50 J/molK at room temperature over the Dulong-Petit limit reported by Bin Chen \textit{et al.}\cite{BinChen2013} is far too large to be reconciled with our findings.

\subsection{2dim Ferromagnetic Magnon Heat Capacity}

We finally comment on the linear contribution  to the low temperature heat capacity which amounts to $\sim$109 mJ/molK$^{2}$, again too large to be attributed to the conduction electron contribution. However, the increase \textit{linear} in temperature, also reported by Bin Chen \textit{et al.} with a similar magnitude, can be  readily reconciled if we take ferromagnetic magnon excitations in a 2dim lattice into account. Adopting a quadratic dispersion for long-wavelength magnons in a 2dim ferromagnet the magnon contribution to the heat capacity at low temperatures varies linearly with temperature.\cite{Miedema2001}
From inelastic neutron scattering data, Song Bao \textit{et al.} have derived spin wave stiffness constants for FGT between 57 and 69 meV\AA$^2$ for the in-plane spin waves.\cite{SongBao2022}
For the magnon heat capacity, $C_{\rm mag }$, of a 2dim honeycomb lattice, Grosu \textit{et al.} arrived at\cite{Grosu2020}
\begin{equation}
C_{\rm {mag}} = R \frac{{3\sqrt{3}\,\,\pi^2} a^2}{24D}  T,
 \label{Magnon}
\end{equation}
where $R$ is the molar gas constant, $a\approx$ 4\AA\ the in-plane lattice parameter and $D$ the spin wave stiffness constant.
With $D~\approx$~63~meV\AA$^2$ one estimates a magnon contribution linear in temperature of
\begin{equation*}
C_{\rm {mag}}/T = 123~\rm{mJ/molK^2}.
 \label{MagnonCp}
\end{equation*}
\RKK{Adding the conduction electron contribution of $\sim$14~mJ/molK$^2$,  this finding is  in fair agreement with our experimental observation of 109 mJ/molK$^2$ found from the Sommerfeld plot at low temperatures (see Fig. \ref{Fig-2}).}

\section{Summary and Conclusion}
Using temperature dependent XRPD  measurements we  have determined the hexagonal lattice parameters of the 2dim van der Waals ferromagnet Fe$_{2.92(1)}$Ge$_{1.02(3)}$Te$_2$
which exhibits a Curie temperature of $\sim$~217~K.
Spontaneous magnetostriction induced by the transition into the  ferromagnetic state \RKK{leads to an expansion of the vdW coupled layers, whereas a noticeable effect of the spontaneous magnetostriction perpendicular to the layers is not seen in the $c$-lattice parameters}. For $T \rightarrow$ 0 K, the in-plane spontaneous magnetostriction \RKK{can be extrapolated to} $\sim$-220$\times$10$^{-6}$~K$^{-1}$ by more than a factor of two smaller than the
\textit{ab-initio} calculations Zhuang \textit{et al.}.

Using the linear volume thermal expansion coefficient in the paramagnetic regime we estimate the difference of the specific heats determined at constant pressure and constant volume. The \RKK{majority of the} linear Sommerfeld-type contribution to the low-temperature heat capacity can be quantitatively ascribed to 2dim ferromagnetic magnon excitations.

In summary, our experimental results evidence a substantial in-plane magnetostriction for a sample of FGT with composition Fe$_{2.92(1)}$Ge$_{1.02(3)}$Te$_2$ having a Curie temperature of 217~K.
\RKK{}
A careful reconsideration of the heat capacities of such sample enables us to correct and understand previously reported  linear contributions to the low- and high-temperature heat capacity data of FGT. Particularly noteworthy is the low-temperature magnon heat capacity contribution, linear in temperature, which can be quantitatively attributed to the layered ferromagnetic character of FGT.

\begin{acknowledgments}
\textit{Acknowledgments}
We thank V. Duppel for performing the SEM/EDS measurements and Hj. Deiseroth for valuable suggestions as to the preparation of FGT. A useful communication with Houlong L. Zhuang is gratefully acknowledged.
\end{acknowledgments}

\bibliography{biblio}

\begin{thebibliography}{38}%
\makeatletter
\providecommand \@ifxundefined [1]{%
 \@ifx{#1\undefined}
}%
\providecommand \@ifnum [1]{%
 \ifnum #1\expandafter \@firstoftwo
 \else \expandafter \@secondoftwo
 \fi
}%
\providecommand \@ifx [1]{%
 \ifx #1\expandafter \@firstoftwo
 \else \expandafter \@secondoftwo
 \fi
}%
\providecommand \natexlab [1]{#1}%
\providecommand \enquote  [1]{``#1''}%
\providecommand \bibnamefont  [1]{#1}%
\providecommand \bibfnamefont [1]{#1}%
\providecommand \citenamefont [1]{#1}%
\providecommand \href@noop [0]{\@secondoftwo}%
\providecommand \href [0]{\begingroup \@sanitize@url \@href}%
\providecommand \@href[1]{\@@startlink{#1}\@@href}%
\providecommand \@@href[1]{\endgroup#1\@@endlink}%
\providecommand \@sanitize@url [0]{\catcode `\\12\catcode `\$12\catcode
  `\&12\catcode `\#12\catcode `\^12\catcode `\_12\catcode `\%12\relax}%
\providecommand \@@startlink[1]{}%
\providecommand \@@endlink[0]{}%
\providecommand \url  [0]{\begingroup\@sanitize@url \@url }%
\providecommand \@url [1]{\endgroup\@href {#1}{\urlprefix }}%
\providecommand \urlprefix  [0]{URL }%
\providecommand \Eprint [0]{\href }%
\providecommand \doibase [0]{https://doi.org/}%
\providecommand \selectlanguage [0]{\@gobble}%
\providecommand \bibinfo  [0]{\@secondoftwo}%
\providecommand \bibfield  [0]{\@secondoftwo}%
\providecommand \translation [1]{[#1]}%
\providecommand \BibitemOpen [0]{}%
\providecommand \bibitemStop [0]{}%
\providecommand \bibitemNoStop [0]{.\EOS\space}%
\providecommand \EOS [0]{\spacefactor3000\relax}%
\providecommand \BibitemShut  [1]{\csname bibitem#1\endcsname}%
\let\auto@bib@innerbib\@empty
\bibitem [{\citenamefont {Deiseroth}\ \emph {et~al.}(2006)\citenamefont
  {Deiseroth}, \citenamefont {Aleksandrov}, \citenamefont {Reiner},
  \citenamefont {Kienle},\ and\ \citenamefont {Kremer}}]{Deiseroth2006}%
  \BibitemOpen
  \bibfield  {author} {\bibinfo {author} {\bibfnamefont {H.-J.}\ \bibnamefont
  {Deiseroth}}, \bibinfo {author} {\bibfnamefont {K.}~\bibnamefont
  {Aleksandrov}}, \bibinfo {author} {\bibfnamefont {C.}~\bibnamefont {Reiner}},
  \bibinfo {author} {\bibfnamefont {L.}~\bibnamefont {Kienle}},\ and\ \bibinfo
  {author} {\bibfnamefont {R.~K.}\ \bibnamefont {Kremer}},\ }\bibfield  {title}
  {\bibinfo {title} {$\mathrm{Fe_{3}GeTe_2}$ and $\mathrm{Ni_{3}GeTe_2}$ –
  two new layered transition-metal compounds: Crystal structures, hrtem
  investigations, and magnetic and electrical properties},\ }\href
  {https://doi.org/https://doi.org/10.1002/ejic.200501020} {\bibfield
  {journal} {\bibinfo  {journal} {Eur. J. Inorg. Chem.}\ }\textbf {\bibinfo
  {volume} {2006}},\ \bibinfo {pages} {1561} (\bibinfo {year}
  {2006})}\BibitemShut {NoStop}%
\bibitem [{\citenamefont {Chen}\ \emph {et~al.}(2013)\citenamefont {Chen},
  \citenamefont {Yang}, \citenamefont {Wang}, \citenamefont {Imai},
  \citenamefont {Ohta}, \citenamefont {Michioka}, \citenamefont {Yoshimura},\
  and\ \citenamefont {Fang}}]{BinChen2013}%
  \BibitemOpen
  \bibfield  {author} {\bibinfo {author} {\bibfnamefont {B.}~\bibnamefont
  {Chen}}, \bibinfo {author} {\bibfnamefont {J.}~\bibnamefont {Yang}}, \bibinfo
  {author} {\bibfnamefont {H.}~\bibnamefont {Wang}}, \bibinfo {author}
  {\bibfnamefont {M.}~\bibnamefont {Imai}}, \bibinfo {author} {\bibfnamefont
  {H.}~\bibnamefont {Ohta}}, \bibinfo {author} {\bibfnamefont {C.}~\bibnamefont
  {Michioka}}, \bibinfo {author} {\bibfnamefont {K.}~\bibnamefont
  {Yoshimura}},\ and\ \bibinfo {author} {\bibfnamefont {M.}~\bibnamefont
  {Fang}},\ }\bibfield  {title} {\bibinfo {title} {Magnetic properties of
  layered itinerant electron ferromagnet $\mathrm{Fe_{3}GeTe_2}$},\ }\href
  {https://doi.org/10.7566/JPSJ.82.124711} {\bibfield  {journal} {\bibinfo
  {journal} {J. Phys. Soc. Jpn.}\ }\textbf {\bibinfo {volume} {82}},\ \bibinfo
  {pages} {124711} (\bibinfo {year} {2013})},\ \Eprint
  {https://arxiv.org/abs/https://doi.org/10.7566/JPSJ.82.124711}
  {https://doi.org/10.7566/JPSJ.82.124711} \BibitemShut {NoStop}%
\bibitem [{\citenamefont {Yi}\ \emph {et~al.}(2017)\citenamefont {Yi},
  \citenamefont {Zhuang}, \citenamefont {Zou}, \citenamefont {Wu},
  \citenamefont {Cao}, \citenamefont {Tang}, \citenamefont {Calder},
  \citenamefont {Kent}, \citenamefont {Mandrus},\ and\ \citenamefont
  {Gai}}]{Yi2017}%
  \BibitemOpen
  \bibfield  {author} {\bibinfo {author} {\bibfnamefont {J.}~\bibnamefont
  {Yi}}, \bibinfo {author} {\bibfnamefont {H.}~\bibnamefont {Zhuang}}, \bibinfo
  {author} {\bibfnamefont {Q.}~\bibnamefont {Zou}}, \bibinfo {author}
  {\bibfnamefont {Z.}~\bibnamefont {Wu}}, \bibinfo {author} {\bibfnamefont
  {G.}~\bibnamefont {Cao}}, \bibinfo {author} {\bibfnamefont {S.}~\bibnamefont
  {Tang}}, \bibinfo {author} {\bibfnamefont {S.~A.}\ \bibnamefont {Calder}},
  \bibinfo {author} {\bibfnamefont {P.~R.~C.}\ \bibnamefont {Kent}}, \bibinfo
  {author} {\bibfnamefont {D.}~\bibnamefont {Mandrus}},\ and\ \bibinfo {author}
  {\bibfnamefont {Z.}~\bibnamefont {Gai}},\ }\bibfield  {title} {\bibinfo
  {title} {Competing antiferromagnetism in a quasi-2d itinerant ferromagnet:
  $\mathrm{Fe_{3}GeTe_2}$},\ }\bibfield  {journal} {\bibinfo  {journal} {2D
  Materials}\ }\textbf {\bibinfo {volume} {4}},\ \href
  {https://doi.org/10.1088/2053-1583/4/1/011005} {10.1088/2053-1583/4/1/011005}
  (\bibinfo {year} {2017})\BibitemShut {NoStop}%
\bibitem [{\citenamefont {Jang}\ \emph {et~al.}(2020)\citenamefont {Jang},
  \citenamefont {Yoon}, \citenamefont {Jeong}, \citenamefont {Ryee},
  \citenamefont {Kim},\ and\ \citenamefont {Han}}]{Jang2020}%
  \BibitemOpen
  \bibfield  {author} {\bibinfo {author} {\bibfnamefont {S.~W.}\ \bibnamefont
  {Jang}}, \bibinfo {author} {\bibfnamefont {H.}~\bibnamefont {Yoon}}, \bibinfo
  {author} {\bibfnamefont {M.~Y.}\ \bibnamefont {Jeong}}, \bibinfo {author}
  {\bibfnamefont {S.}~\bibnamefont {Ryee}}, \bibinfo {author} {\bibfnamefont
  {H.-S.}\ \bibnamefont {Kim}},\ and\ \bibinfo {author} {\bibfnamefont {M.~J.}\
  \bibnamefont {Han}},\ }\bibfield  {title} {\bibinfo {title} {Origin of
  ferromagnetism and the effect of doping on $\mathrm{Fe_{3}GeTe_2}$},\ }\href
  {https://doi.org/10.1039/c9nr10171c} {\bibfield  {journal} {\bibinfo
  {journal} {Nanoscale}\ }\textbf {\bibinfo {volume} {12}},\ \bibinfo {pages}
  {13501} (\bibinfo {year} {2020})}\BibitemShut {NoStop}%
\bibitem [{\citenamefont {Tan}\ \emph {et~al.}(2018)\citenamefont {Tan},
  \citenamefont {Lee}, \citenamefont {Jung}, \citenamefont {Park},
  \citenamefont {Albarakati}, \citenamefont {Partridge}, \citenamefont {Field},
  \citenamefont {McCulloch}, \citenamefont {Wang},\ and\ \citenamefont
  {Lee}}]{Tan2018}%
  \BibitemOpen
  \bibfield  {author} {\bibinfo {author} {\bibfnamefont {C.}~\bibnamefont
  {Tan}}, \bibinfo {author} {\bibfnamefont {J.}~\bibnamefont {Lee}}, \bibinfo
  {author} {\bibfnamefont {S.-G.}\ \bibnamefont {Jung}}, \bibinfo {author}
  {\bibfnamefont {T.}~\bibnamefont {Park}}, \bibinfo {author} {\bibfnamefont
  {S.}~\bibnamefont {Albarakati}}, \bibinfo {author} {\bibfnamefont
  {J.}~\bibnamefont {Partridge}}, \bibinfo {author} {\bibfnamefont {M.~R.}\
  \bibnamefont {Field}}, \bibinfo {author} {\bibfnamefont {D.~G.}\ \bibnamefont
  {McCulloch}}, \bibinfo {author} {\bibfnamefont {L.}~\bibnamefont {Wang}},\
  and\ \bibinfo {author} {\bibfnamefont {C.}~\bibnamefont {Lee}},\ }\bibfield
  {title} {\bibinfo {title} {Hard magnetic properties in nanoflake van der
  waals $\mathrm{Fe_{3}GeTe_2}$},\ }\bibfield  {journal} {\bibinfo  {journal}
  {Nat. Commun.}\ }\textbf {\bibinfo {volume} {9}},\ \href
  {https://doi.org/10.1038/s41467-018-04018-w} {10.1038/s41467-018-04018-w}
  (\bibinfo {year} {2018})\BibitemShut {NoStop}%
\bibitem [{\citenamefont {Wang}\ \emph {et~al.}(2018)\citenamefont {Wang},
  \citenamefont {Sapkota}, \citenamefont {Taniguchi}, \citenamefont {Watanabe},
  \citenamefont {Mandrus},\ and\ \citenamefont {Morpurgo}}]{Wang2018}%
  \BibitemOpen
  \bibfield  {author} {\bibinfo {author} {\bibfnamefont {Z.}~\bibnamefont
  {Wang}}, \bibinfo {author} {\bibfnamefont {D.}~\bibnamefont {Sapkota}},
  \bibinfo {author} {\bibfnamefont {T.}~\bibnamefont {Taniguchi}}, \bibinfo
  {author} {\bibfnamefont {K.}~\bibnamefont {Watanabe}}, \bibinfo {author}
  {\bibfnamefont {D.}~\bibnamefont {Mandrus}},\ and\ \bibinfo {author}
  {\bibfnamefont {A.~F.}\ \bibnamefont {Morpurgo}},\ }\bibfield  {title}
  {\bibinfo {title} {Tunneling spin valves based on
  $\mathrm{Fe_{3}GeTe_2}$/hbn/$\mathrm{Fe_{3}GeTe_2}$ van der {W}aals
  heterostructures},\ }\href {https://doi.org/10.1021/acs.nanolett.8b01278}
  {\bibfield  {journal} {\bibinfo  {journal} {Nano Letters}\ }\textbf {\bibinfo
  {volume} {18}},\ \bibinfo {pages} {4303} (\bibinfo {year}
  {2018})}\BibitemShut {NoStop}%
\bibitem [{\citenamefont {Kim}\ \emph {et~al.}(2018)\citenamefont {Kim},
  \citenamefont {Seo}, \citenamefont {Lee}, \citenamefont {Ko}, \citenamefont
  {Kim}, \citenamefont {Jang}, \citenamefont {Ok}, \citenamefont {Lee},
  \citenamefont {Jo}, \citenamefont {Kang}, \citenamefont {Shim}, \citenamefont
  {Kim}, \citenamefont {Yeom}, \citenamefont {Il~Min},\ and\ \citenamefont
  {Yang}}]{Kim2018}%
  \BibitemOpen
  \bibfield  {author} {\bibinfo {author} {\bibfnamefont {K.}~\bibnamefont
  {Kim}}, \bibinfo {author} {\bibfnamefont {J.}~\bibnamefont {Seo}}, \bibinfo
  {author} {\bibfnamefont {E.}~\bibnamefont {Lee}}, \bibinfo {author}
  {\bibfnamefont {K.-T.}\ \bibnamefont {Ko}}, \bibinfo {author} {\bibfnamefont
  {B.~S.}\ \bibnamefont {Kim}}, \bibinfo {author} {\bibfnamefont {B.~G.}\
  \bibnamefont {Jang}}, \bibinfo {author} {\bibfnamefont {J.~M.}\ \bibnamefont
  {Ok}}, \bibinfo {author} {\bibfnamefont {J.}~\bibnamefont {Lee}}, \bibinfo
  {author} {\bibfnamefont {Y.~J.}\ \bibnamefont {Jo}}, \bibinfo {author}
  {\bibfnamefont {W.}~\bibnamefont {Kang}}, \bibinfo {author} {\bibfnamefont
  {J.~H.}\ \bibnamefont {Shim}}, \bibinfo {author} {\bibfnamefont
  {C.}~\bibnamefont {Kim}}, \bibinfo {author} {\bibfnamefont {H.~W.}\
  \bibnamefont {Yeom}}, \bibinfo {author} {\bibfnamefont {B.}~\bibnamefont
  {Il~Min}},\ and\ \bibinfo {author} {\bibfnamefont {J.~S.}\ \bibnamefont
  {Yang}, \bibfnamefont {Bohm-Jung~Kim}},\ }\bibfield  {title} {\bibinfo
  {title} {Large anomalous hall current induced by topological nodal lines in a
  ferromagnetic van der waals semimetal},\ }\href
  {https://doi.org/doi.org/10.1038/s41563-018-0132-3} {\bibfield  {journal}
  {\bibinfo  {journal} {Nature Materials}\ }\textbf {\bibinfo {volume} {17}},\
  \bibinfo {pages} {794} (\bibinfo {year} {2018})}\BibitemShut {NoStop}%
\bibitem [{\citenamefont {Liu}\ \emph {et~al.}(2018)\citenamefont {Liu},
  \citenamefont {Stavitski}, \citenamefont {Attenkofer},\ and\ \citenamefont
  {Petrovic}}]{Liu2018}%
  \BibitemOpen
  \bibfield  {author} {\bibinfo {author} {\bibfnamefont {Y.}~\bibnamefont
  {Liu}}, \bibinfo {author} {\bibfnamefont {E.}~\bibnamefont {Stavitski}},
  \bibinfo {author} {\bibfnamefont {K.}~\bibnamefont {Attenkofer}},\ and\
  \bibinfo {author} {\bibfnamefont {C.}~\bibnamefont {Petrovic}},\ }\bibfield
  {title} {\bibinfo {title} {Anomalous hall effect in the van der waals bonded
  ferromagnet $\mathrm{Fe_{3-x}GeTe_2}$},\ }\href
  {https://doi.org/10.1103/PhysRevB.97.165415} {\bibfield  {journal} {\bibinfo
  {journal} {Phys. Rev. B}\ }\textbf {\bibinfo {volume} {97}},\ \bibinfo
  {pages} {165415} (\bibinfo {year} {2018})}\BibitemShut {NoStop}%
\bibitem [{\citenamefont {Zhuang}\ \emph {et~al.}(2016)\citenamefont {Zhuang},
  \citenamefont {Kent},\ and\ \citenamefont {Hennig}}]{Zhuang2016}%
  \BibitemOpen
  \bibfield  {author} {\bibinfo {author} {\bibfnamefont {H.~L.}\ \bibnamefont
  {Zhuang}}, \bibinfo {author} {\bibfnamefont {P.~R.~C.}\ \bibnamefont
  {Kent}},\ and\ \bibinfo {author} {\bibfnamefont {R.~G.}\ \bibnamefont
  {Hennig}},\ }\bibfield  {title} {\bibinfo {title} {Strong anisotropy and
  magnetostriction in the two-dimensional stoner ferromagnet
  $\mathrm{Fe_{3}GeTe_2}$},\ }\href
  {https://doi.org/10.1103/PhysRevB.93.134407} {\bibfield  {journal} {\bibinfo
  {journal} {Phys. Rev. B}\ }\textbf {\bibinfo {volume} {93}},\ \bibinfo
  {pages} {134407} (\bibinfo {year} {2016})}\BibitemShut {NoStop}%
\bibitem [{\citenamefont {Fei}\ \emph {et~al.}(2018)\citenamefont {Fei},
  \citenamefont {Huang}, \citenamefont {Malinowski}, \citenamefont {Wang},
  \citenamefont {Song}, \citenamefont {Sanchez}, \citenamefont {Yao},
  \citenamefont {Xiao}, \citenamefont {Zhu}, \citenamefont {May}, \citenamefont
  {Wu}, \citenamefont {Cobden}, \citenamefont {Chu},\ and\ \citenamefont
  {Xu}}]{Fei2018}%
  \BibitemOpen
  \bibfield  {author} {\bibinfo {author} {\bibfnamefont {Z.}~\bibnamefont
  {Fei}}, \bibinfo {author} {\bibfnamefont {B.}~\bibnamefont {Huang}}, \bibinfo
  {author} {\bibfnamefont {P.}~\bibnamefont {Malinowski}}, \bibinfo {author}
  {\bibfnamefont {W.}~\bibnamefont {Wang}}, \bibinfo {author} {\bibfnamefont
  {T.}~\bibnamefont {Song}}, \bibinfo {author} {\bibfnamefont {J.}~\bibnamefont
  {Sanchez}}, \bibinfo {author} {\bibfnamefont {W.}~\bibnamefont {Yao}},
  \bibinfo {author} {\bibfnamefont {D.}~\bibnamefont {Xiao}}, \bibinfo {author}
  {\bibfnamefont {X.}~\bibnamefont {Zhu}}, \bibinfo {author} {\bibfnamefont
  {A.~F.}\ \bibnamefont {May}}, \bibinfo {author} {\bibfnamefont
  {W.}~\bibnamefont {Wu}}, \bibinfo {author} {\bibfnamefont {D.~H.}\
  \bibnamefont {Cobden}}, \bibinfo {author} {\bibfnamefont {J.-H.}\
  \bibnamefont {Chu}},\ and\ \bibinfo {author} {\bibfnamefont {X.}~\bibnamefont
  {Xu}},\ }\bibfield  {title} {\bibinfo {title} {Two-dimensional itinerant
  ferromagnetism in atomically thin $\mathrm{Fe_{3}GeTe_2}$},\ }\href
  {https://doi.org/10.1038/s41563-018-0149-7} {\bibfield  {journal} {\bibinfo
  {journal} {Nature Materials}\ }\textbf {\bibinfo {volume} {17}},\ \bibinfo
  {pages} {778} (\bibinfo {year} {2018})}\BibitemShut {NoStop}%
\bibitem [{\citenamefont {Deng}\ \emph {et~al.}(2018)\citenamefont {Deng},
  \citenamefont {Yu}, \citenamefont {Song}, \citenamefont {Zhang},
  \citenamefont {Wang}, \citenamefont {Sun}, \citenamefont {Yi}, \citenamefont
  {Wu}, \citenamefont {Wu}, \citenamefont {Zhu}, \citenamefont {Wang},
  \citenamefont {Chen},\ and\ \citenamefont {Zhang}}]{Deng2018}%
  \BibitemOpen
  \bibfield  {author} {\bibinfo {author} {\bibfnamefont {Y.}~\bibnamefont
  {Deng}}, \bibinfo {author} {\bibfnamefont {Y.}~\bibnamefont {Yu}}, \bibinfo
  {author} {\bibfnamefont {Y.}~\bibnamefont {Song}}, \bibinfo {author}
  {\bibfnamefont {J.}~\bibnamefont {Zhang}}, \bibinfo {author} {\bibfnamefont
  {N.~Z.}\ \bibnamefont {Wang}}, \bibinfo {author} {\bibfnamefont
  {Z.}~\bibnamefont {Sun}}, \bibinfo {author} {\bibfnamefont {Y.}~\bibnamefont
  {Yi}}, \bibinfo {author} {\bibfnamefont {Y.~Z.}\ \bibnamefont {Wu}}, \bibinfo
  {author} {\bibfnamefont {S.}~\bibnamefont {Wu}}, \bibinfo {author}
  {\bibfnamefont {J.}~\bibnamefont {Zhu}}, \bibinfo {author} {\bibfnamefont
  {J.}~\bibnamefont {Wang}}, \bibinfo {author} {\bibfnamefont {X.~H.}\
  \bibnamefont {Chen}},\ and\ \bibinfo {author} {\bibfnamefont
  {Y.}~\bibnamefont {Zhang}},\ }\bibfield  {title} {\bibinfo {title}
  {Gate-tunable room-temperature ferromagnetism in two-dimensional
  $\mathrm{Fe_{3}GeTe_2}$},\ }\href
  {https://doi.org/doi.org/10.1038/s41586-018-0626-9} {\bibfield  {journal}
  {\bibinfo  {journal} {Nature}\ }\textbf {\bibinfo {volume} {563}},\ \bibinfo
  {pages} {94} (\bibinfo {year} {2018})}\BibitemShut {NoStop}%
\bibitem [{\citenamefont {Merte}\ \emph {et~al.}(2021)\citenamefont {Merte},
  \citenamefont {Freimuth}, \citenamefont {Adamantopoulos}, \citenamefont {Go},
  \citenamefont {Saunderson}, \citenamefont {Kl\"aui}, \citenamefont
  {Plucinski}, \citenamefont {Gomonay}, \citenamefont {Bl\"ugel},\ and\
  \citenamefont {Mokrousov}}]{Merte2021}%
  \BibitemOpen
  \bibfield  {author} {\bibinfo {author} {\bibfnamefont {M.}~\bibnamefont
  {Merte}}, \bibinfo {author} {\bibfnamefont {F.}~\bibnamefont {Freimuth}},
  \bibinfo {author} {\bibfnamefont {T.}~\bibnamefont {Adamantopoulos}},
  \bibinfo {author} {\bibfnamefont {D.}~\bibnamefont {Go}}, \bibinfo {author}
  {\bibfnamefont {T.~G.}\ \bibnamefont {Saunderson}}, \bibinfo {author}
  {\bibfnamefont {M.}~\bibnamefont {Kl\"aui}}, \bibinfo {author} {\bibfnamefont
  {L.}~\bibnamefont {Plucinski}}, \bibinfo {author} {\bibfnamefont
  {O.}~\bibnamefont {Gomonay}}, \bibinfo {author} {\bibfnamefont
  {S.}~\bibnamefont {Bl\"ugel}},\ and\ \bibinfo {author} {\bibfnamefont
  {Y.}~\bibnamefont {Mokrousov}},\ }\bibfield  {title} {\bibinfo {title}
  {Photocurrents of charge and spin in monolayer $\mathrm{Fe_{3}GeTe_2}$},\
  }\href {https://doi.org/10.1103/PhysRevB.104.L220405} {\bibfield  {journal}
  {\bibinfo  {journal} {Phys. Rev. B}\ }\textbf {\bibinfo {volume} {104}},\
  \bibinfo {pages} {L220405} (\bibinfo {year} {2021})}\BibitemShut {NoStop}%
\bibitem [{\citenamefont {Ding}\ \emph {et~al.}(2020)\citenamefont {Ding},
  \citenamefont {Li}, \citenamefont {Xu}, \citenamefont {Li}, \citenamefont
  {Hou}, \citenamefont {Liu}, \citenamefont {Xi}, \citenamefont {Xu},
  \citenamefont {Yao},\ and\ \citenamefont {Wang}}]{Ding2020}%
  \BibitemOpen
  \bibfield  {author} {\bibinfo {author} {\bibfnamefont {B.}~\bibnamefont
  {Ding}}, \bibinfo {author} {\bibfnamefont {Z.}~\bibnamefont {Li}}, \bibinfo
  {author} {\bibfnamefont {G.}~\bibnamefont {Xu}}, \bibinfo {author}
  {\bibfnamefont {H.}~\bibnamefont {Li}}, \bibinfo {author} {\bibfnamefont
  {Z.}~\bibnamefont {Hou}}, \bibinfo {author} {\bibfnamefont {E.}~\bibnamefont
  {Liu}}, \bibinfo {author} {\bibfnamefont {X.}~\bibnamefont {Xi}}, \bibinfo
  {author} {\bibfnamefont {F.}~\bibnamefont {Xu}}, \bibinfo {author}
  {\bibfnamefont {Y.}~\bibnamefont {Yao}},\ and\ \bibinfo {author}
  {\bibfnamefont {W.}~\bibnamefont {Wang}},\ }\bibfield  {title} {\bibinfo
  {title} {Observation of magnetic skyrmion bubbles in a van der waals
  ferromagnet $\mathrm{Fe_3GeTe_2}$},\ }\href
  {https://doi.org/10.1021/acs.nanolett.9b03453} {\bibfield  {journal}
  {\bibinfo  {journal} {Nano Letters}\ }\textbf {\bibinfo {volume} {20}},\
  \bibinfo {pages} {868} (\bibinfo {year} {2020})},\ \bibinfo {note} {pMID:
  31869236},\ \Eprint
  {https://arxiv.org/abs/https://doi.org/10.1021/acs.nanolett.9b03453}
  {https://doi.org/10.1021/acs.nanolett.9b03453} \BibitemShut {NoStop}%
\bibitem [{\citenamefont {Meijer}\ \emph {et~al.}(2020)\citenamefont {Meijer},
  \citenamefont {Lucassen}, \citenamefont {Duine}, \citenamefont {Swagten},
  \citenamefont {Koopmans}, \citenamefont {Lavrijsen},\ and\ \citenamefont
  {Guimarães}}]{Meijer2020}%
  \BibitemOpen
  \bibfield  {author} {\bibinfo {author} {\bibfnamefont {M.~J.}\ \bibnamefont
  {Meijer}}, \bibinfo {author} {\bibfnamefont {J.}~\bibnamefont {Lucassen}},
  \bibinfo {author} {\bibfnamefont {R.~A.}\ \bibnamefont {Duine}}, \bibinfo
  {author} {\bibfnamefont {H.~J.}\ \bibnamefont {Swagten}}, \bibinfo {author}
  {\bibfnamefont {B.}~\bibnamefont {Koopmans}}, \bibinfo {author}
  {\bibfnamefont {R.}~\bibnamefont {Lavrijsen}},\ and\ \bibinfo {author}
  {\bibfnamefont {M.~H.~D.}\ \bibnamefont {Guimarães}},\ }\bibfield  {title}
  {\bibinfo {title} {Chiral spin spirals at the surface of the van der waals
  ferromagnet fe3gete2},\ }\href {https://doi.org/10.1021/acs.nanolett.0c03111}
  {\bibfield  {journal} {\bibinfo  {journal} {Nano Letters}\ }\textbf {\bibinfo
  {volume} {20}},\ \bibinfo {pages} {8563} (\bibinfo {year} {2020})},\ \bibinfo
  {note} {pMID: 33238096},\ \Eprint
  {https://arxiv.org/abs/https://doi.org/10.1021/acs.nanolett.0c03111}
  {https://doi.org/10.1021/acs.nanolett.0c03111} \BibitemShut {NoStop}%
\bibitem [{\citenamefont {Wu}\ \emph {et~al.}(2020)\citenamefont {Wu},
  \citenamefont {Zhang}, \citenamefont {Zhang}, \citenamefont {Wang},
  \citenamefont {Zhu}, \citenamefont {Hu}, \citenamefont {Yin}, \citenamefont
  {Wong}, \citenamefont {Fang}, \citenamefont {Wan}, \citenamefont {Han},
  \citenamefont {Shao}, \citenamefont {Taniguchi}, \citenamefont {Watanabe},
  \citenamefont {Zang}, \citenamefont {Mao}, \citenamefont {Zhang},\ and\
  \citenamefont {Wang}}]{Wu2020}%
  \BibitemOpen
  \bibfield  {author} {\bibinfo {author} {\bibfnamefont {Y.}~\bibnamefont
  {Wu}}, \bibinfo {author} {\bibfnamefont {S.}~\bibnamefont {Zhang}}, \bibinfo
  {author} {\bibfnamefont {J.}~\bibnamefont {Zhang}}, \bibinfo {author}
  {\bibfnamefont {W.}~\bibnamefont {Wang}}, \bibinfo {author} {\bibfnamefont
  {Y.~L.}\ \bibnamefont {Zhu}}, \bibinfo {author} {\bibfnamefont
  {J.}~\bibnamefont {Hu}}, \bibinfo {author} {\bibfnamefont {G.}~\bibnamefont
  {Yin}}, \bibinfo {author} {\bibfnamefont {K.}~\bibnamefont {Wong}}, \bibinfo
  {author} {\bibfnamefont {C.}~\bibnamefont {Fang}}, \bibinfo {author}
  {\bibfnamefont {C.}~\bibnamefont {Wan}}, \bibinfo {author} {\bibfnamefont
  {X.}~\bibnamefont {Han}}, \bibinfo {author} {\bibfnamefont {Q.}~\bibnamefont
  {Shao}}, \bibinfo {author} {\bibfnamefont {T.}~\bibnamefont {Taniguchi}},
  \bibinfo {author} {\bibfnamefont {K.}~\bibnamefont {Watanabe}}, \bibinfo
  {author} {\bibfnamefont {J.}~\bibnamefont {Zang}}, \bibinfo {author}
  {\bibfnamefont {Z.}~\bibnamefont {Mao}}, \bibinfo {author} {\bibfnamefont
  {X.}~\bibnamefont {Zhang}},\ and\ \bibinfo {author} {\bibfnamefont {K.~L.}\
  \bibnamefont {Wang}},\ }\bibfield  {title} {\bibinfo {title} {N\'eel-type
  skyrmion in $\mathrm{WTe_2}$/$\mathrm{Fe_{3}GeTe_2}$ van der waals
  heterostructure},\ }\bibfield  {journal} {\bibinfo  {journal} {Nat. Commun.}\
  }\textbf {\bibinfo {volume} {11}},\ \href
  {https://doi.org/10.1038/s41467-020-17566-x} {10.1038/s41467-020-17566-x}
  (\bibinfo {year} {2020})\BibitemShut {NoStop}%
\bibitem [{\citenamefont {Yang}\ \emph {et~al.}(2020)\citenamefont {Yang},
  \citenamefont {Li}, \citenamefont {Chopdekar}, \citenamefont {Dhall},
  \citenamefont {Turner}, \citenamefont {Carlstr\"om}, \citenamefont {Ophus},
  \citenamefont {Klewe}, \citenamefont {Shafer}, \citenamefont {N'Diaye},
  \citenamefont {Choi}, \citenamefont {Chen}, \citenamefont {Wu}, \citenamefont
  {Hwang}, \citenamefont {Wang},\ and\ \citenamefont {Qiu}}]{Yang2020}%
  \BibitemOpen
  \bibfield  {author} {\bibinfo {author} {\bibfnamefont {M.}~\bibnamefont
  {Yang}}, \bibinfo {author} {\bibfnamefont {Q.}~\bibnamefont {Li}}, \bibinfo
  {author} {\bibfnamefont {V.~R.}\ \bibnamefont {Chopdekar}}, \bibinfo {author}
  {\bibfnamefont {R.}~\bibnamefont {Dhall}}, \bibinfo {author} {\bibfnamefont
  {J.}~\bibnamefont {Turner}}, \bibinfo {author} {\bibfnamefont {J.~D.}\
  \bibnamefont {Carlstr\"om}}, \bibinfo {author} {\bibfnamefont
  {C.}~\bibnamefont {Ophus}}, \bibinfo {author} {\bibfnamefont
  {C.}~\bibnamefont {Klewe}}, \bibinfo {author} {\bibfnamefont
  {P.}~\bibnamefont {Shafer}}, \bibinfo {author} {\bibfnamefont {A.~T.}\
  \bibnamefont {N'Diaye}}, \bibinfo {author} {\bibfnamefont {J.~W.}\
  \bibnamefont {Choi}}, \bibinfo {author} {\bibfnamefont {G.}~\bibnamefont
  {Chen}}, \bibinfo {author} {\bibfnamefont {Y.~Z.}\ \bibnamefont {Wu}},
  \bibinfo {author} {\bibfnamefont {C.}~\bibnamefont {Hwang}}, \bibinfo
  {author} {\bibfnamefont {F.}~\bibnamefont {Wang}},\ and\ \bibinfo {author}
  {\bibfnamefont {Z.~Q.}\ \bibnamefont {Qiu}},\ }\bibfield  {title} {\bibinfo
  {title} {Creation of skyrmions in van der waals ferromagnet
  $\mathrm{Fe_{3}GeTe_2}$ on (co/pd)(n) superlattice},\ }\bibfield  {journal}
  {\bibinfo  {journal} {Sci. Adv.}\ }\textbf {\bibinfo {volume} {6}},\ \href
  {https://doi.org/10.1126/sciadv.abb5157} {10.1126/sciadv.abb5157} (\bibinfo
  {year} {2020})\BibitemShut {NoStop}%
\bibitem [{\citenamefont {Park}\ \emph {et~al.}(2021)\citenamefont {Park},
  \citenamefont {Peng}, \citenamefont {Liang}, \citenamefont {Hallal},
  \citenamefont {Yasin}, \citenamefont {Zhang}, \citenamefont {Song},
  \citenamefont {Kim}, \citenamefont {Kim}, \citenamefont {Weigand},
  \citenamefont {Sch\"utz}, \citenamefont {Finizio}, \citenamefont {Raabe},
  \citenamefont {Garcia}, \citenamefont {Xia}, \citenamefont {Zhou},
  \citenamefont {Ezawa}, \citenamefont {Liu}, \citenamefont {Chang},
  \citenamefont {Koo}, \citenamefont {Kim}, \citenamefont {Chshiev},
  \citenamefont {Fert}, \citenamefont {Yang}, \citenamefont {Yu},\ and\
  \citenamefont {Woo}}]{Park2021}%
  \BibitemOpen
  \bibfield  {author} {\bibinfo {author} {\bibfnamefont {T.-E.}\ \bibnamefont
  {Park}}, \bibinfo {author} {\bibfnamefont {L.}~\bibnamefont {Peng}}, \bibinfo
  {author} {\bibfnamefont {J.}~\bibnamefont {Liang}}, \bibinfo {author}
  {\bibfnamefont {A.}~\bibnamefont {Hallal}}, \bibinfo {author} {\bibfnamefont
  {F.~S.}\ \bibnamefont {Yasin}}, \bibinfo {author} {\bibfnamefont
  {X.}~\bibnamefont {Zhang}}, \bibinfo {author} {\bibfnamefont {K.~M.}\
  \bibnamefont {Song}}, \bibinfo {author} {\bibfnamefont {S.~J.}\ \bibnamefont
  {Kim}}, \bibinfo {author} {\bibfnamefont {K.}~\bibnamefont {Kim}}, \bibinfo
  {author} {\bibfnamefont {M.}~\bibnamefont {Weigand}}, \bibinfo {author}
  {\bibfnamefont {G.}~\bibnamefont {Sch\"utz}}, \bibinfo {author}
  {\bibfnamefont {S.}~\bibnamefont {Finizio}}, \bibinfo {author} {\bibfnamefont
  {J.}~\bibnamefont {Raabe}}, \bibinfo {author} {\bibfnamefont
  {K.}~\bibnamefont {Garcia}}, \bibinfo {author} {\bibfnamefont
  {J.}~\bibnamefont {Xia}}, \bibinfo {author} {\bibfnamefont {Y.}~\bibnamefont
  {Zhou}}, \bibinfo {author} {\bibfnamefont {M.}~\bibnamefont {Ezawa}},
  \bibinfo {author} {\bibfnamefont {X.}~\bibnamefont {Liu}}, \bibinfo {author}
  {\bibfnamefont {J.}~\bibnamefont {Chang}}, \bibinfo {author} {\bibfnamefont
  {H.~C.}\ \bibnamefont {Koo}}, \bibinfo {author} {\bibfnamefont {Y.~D.}\
  \bibnamefont {Kim}}, \bibinfo {author} {\bibfnamefont {M.}~\bibnamefont
  {Chshiev}}, \bibinfo {author} {\bibfnamefont {A.}~\bibnamefont {Fert}},
  \bibinfo {author} {\bibfnamefont {H.}~\bibnamefont {Yang}}, \bibinfo {author}
  {\bibfnamefont {X.}~\bibnamefont {Yu}},\ and\ \bibinfo {author}
  {\bibfnamefont {S.}~\bibnamefont {Woo}},\ }\bibfield  {title} {\bibinfo
  {title} {N\'eel-type skyrmions and their current-induced motion in van der
  waals ferromagnet-based heterostructures},\ }\href
  {https://doi.org/10.1103/PhysRevB.103.104410} {\bibfield  {journal} {\bibinfo
   {journal} {Phys. Rev. B}\ }\textbf {\bibinfo {volume} {103}},\ \bibinfo
  {pages} {104410} (\bibinfo {year} {2021})}\BibitemShut {NoStop}%
\bibitem [{\citenamefont {Xu}\ \emph {et~al.}(2022)\citenamefont {Xu},
  \citenamefont {Li}, \citenamefont {Chen}, \citenamefont {Zhang},
  \citenamefont {Xiang},\ and\ \citenamefont {Bellaiche}}]{Xu2022}%
  \BibitemOpen
  \bibfield  {author} {\bibinfo {author} {\bibfnamefont {C.}~\bibnamefont
  {Xu}}, \bibinfo {author} {\bibfnamefont {X.}~\bibnamefont {Li}}, \bibinfo
  {author} {\bibfnamefont {P.}~\bibnamefont {Chen}}, \bibinfo {author}
  {\bibfnamefont {Y.}~\bibnamefont {Zhang}}, \bibinfo {author} {\bibfnamefont
  {H.}~\bibnamefont {Xiang}},\ and\ \bibinfo {author} {\bibfnamefont
  {L.}~\bibnamefont {Bellaiche}},\ }\bibfield  {title} {\bibinfo {title}
  {Assembling diverse skyrmionic phases in $\mathrm{Fe_{3}GeTe_2}$
  monolayers},\ }\bibfield  {journal} {\bibinfo  {journal} {Adv. Mater.}\
  }\textbf {\bibinfo {volume} {34}},\ \href
  {https://doi.org/10.1002/adma.202107779} {10.1002/adma.202107779} (\bibinfo
  {year} {2022})\BibitemShut {NoStop}%
\bibitem [{\citenamefont {Chen}\ \emph {et~al.}(2022)\citenamefont {Chen},
  \citenamefont {Wang}, \citenamefont {Liu}, \citenamefont {Wang},
  \citenamefont {Wei}, \citenamefont {Fang}, \citenamefont {Wang},
  \citenamefont {Geng}, \citenamefont {Liu}, \citenamefont {Li}, \citenamefont
  {Yu}, \citenamefont {Zhao}, \citenamefont {Miao}, \citenamefont {Li},
  \citenamefont {Wang}, \citenamefont {Nie}, \citenamefont {Zhao},\ and\
  \citenamefont {Wu}}]{Chen2022}%
  \BibitemOpen
  \bibfield  {author} {\bibinfo {author} {\bibfnamefont {X.}~\bibnamefont
  {Chen}}, \bibinfo {author} {\bibfnamefont {H.}~\bibnamefont {Wang}}, \bibinfo
  {author} {\bibfnamefont {H.}~\bibnamefont {Liu}}, \bibinfo {author}
  {\bibfnamefont {C.}~\bibnamefont {Wang}}, \bibinfo {author} {\bibfnamefont
  {G.}~\bibnamefont {Wei}}, \bibinfo {author} {\bibfnamefont {C.}~\bibnamefont
  {Fang}}, \bibinfo {author} {\bibfnamefont {H.}~\bibnamefont {Wang}}, \bibinfo
  {author} {\bibfnamefont {C.}~\bibnamefont {Geng}}, \bibinfo {author}
  {\bibfnamefont {S.}~\bibnamefont {Liu}}, \bibinfo {author} {\bibfnamefont
  {P.}~\bibnamefont {Li}}, \bibinfo {author} {\bibfnamefont {H.}~\bibnamefont
  {Yu}}, \bibinfo {author} {\bibfnamefont {W.}~\bibnamefont {Zhao}}, \bibinfo
  {author} {\bibfnamefont {J.}~\bibnamefont {Miao}}, \bibinfo {author}
  {\bibfnamefont {Y.}~\bibnamefont {Li}}, \bibinfo {author} {\bibfnamefont
  {L.}~\bibnamefont {Wang}}, \bibinfo {author} {\bibfnamefont {T.}~\bibnamefont
  {Nie}}, \bibinfo {author} {\bibfnamefont {J.}~\bibnamefont {Zhao}},\ and\
  \bibinfo {author} {\bibfnamefont {X.}~\bibnamefont {Wu}},\ }\bibfield
  {title} {\bibinfo {title} {Generation and control of terahertz spin currents
  in topology-induced 2d ferromagnetic
  $\mathrm{Fe_{3}GeTe_2}$/$\mathrm{Bi_2Te_3}$ heterostructures},\ }\bibfield
  {journal} {\bibinfo  {journal} {Adv. Mater.}\ }\textbf {\bibinfo {volume}
  {34}},\ \href {https://doi.org/10.1002/adma.202106172}
  {10.1002/adma.202106172} (\bibinfo {year} {2022})\BibitemShut {NoStop}%
\bibitem [{\citenamefont {Milosavljevi\ifmmode~\acute{c}\else \'{c}\fi{}}\
  \emph {et~al.}(2019)\citenamefont {Milosavljevi\ifmmode~\acute{c}\else
  \'{c}\fi{}}, \citenamefont {\ifmmode \check{S}\else
  \v{S}\fi{}olaji\ifmmode~\acute{c}\else \'{c}\fi{}}, \citenamefont
  {Djurdji\ifmmode \acute{c}\else~\'{c}\fi{} Mijin}, \citenamefont {Pe\ifmmode
  \check{s}\else \v{s}\fi{}i\ifmmode~\acute{c}\else \'{c}\fi{}}, \citenamefont
  {Vi\ifmmode \check{s}\else \v{s}\fi{}i\ifmmode~\acute{c}\else \'{c}\fi{}},
  \citenamefont {Liu}, \citenamefont {Petrovic}, \citenamefont
  {Lazarevi\ifmmode~\acute{c}\else \'{c}\fi{}},\ and\ \citenamefont
  {Popovi\ifmmode~\acute{c}\else \'{c}\fi{}}}]{Popovic2019}%
  \BibitemOpen
  \bibfield  {author} {\bibinfo {author} {\bibfnamefont {A.}~\bibnamefont
  {Milosavljevi\ifmmode~\acute{c}\else \'{c}\fi{}}}, \bibinfo {author}
  {\bibfnamefont {A.}~\bibnamefont {\ifmmode \check{S}\else
  \v{S}\fi{}olaji\ifmmode~\acute{c}\else \'{c}\fi{}}}, \bibinfo {author}
  {\bibfnamefont {S.}~\bibnamefont {Djurdji\ifmmode \acute{c}\else~\'{c}\fi{}
  Mijin}}, \bibinfo {author} {\bibfnamefont {J.}~\bibnamefont {Pe\ifmmode
  \check{s}\else \v{s}\fi{}i\ifmmode~\acute{c}\else \'{c}\fi{}}}, \bibinfo
  {author} {\bibfnamefont {B.}~\bibnamefont {Vi\ifmmode \check{s}\else
  \v{s}\fi{}i\ifmmode~\acute{c}\else \'{c}\fi{}}}, \bibinfo {author}
  {\bibfnamefont {Y.}~\bibnamefont {Liu}}, \bibinfo {author} {\bibfnamefont
  {C.}~\bibnamefont {Petrovic}}, \bibinfo {author} {\bibfnamefont
  {N.}~\bibnamefont {Lazarevi\ifmmode~\acute{c}\else \'{c}\fi{}}},\ and\
  \bibinfo {author} {\bibfnamefont {Z.~V.}\ \bibnamefont
  {Popovi\ifmmode~\acute{c}\else \'{c}\fi{}}},\ }\bibfield  {title} {\bibinfo
  {title} {Lattice dynamics and phase transitions in
  $\mathrm{Fe_{3-x}GeTe_2}$},\ }\href
  {https://doi.org/10.1103/PhysRevB.99.214304} {\bibfield  {journal} {\bibinfo
  {journal} {Phys. Rev. B}\ }\textbf {\bibinfo {volume} {99}},\ \bibinfo
  {pages} {214304} (\bibinfo {year} {2019})}\BibitemShut {NoStop}%
\bibitem [{\citenamefont {May}\ \emph {et~al.}(2016)\citenamefont {May},
  \citenamefont {Calder}, \citenamefont {Cantoni}, \citenamefont {Cao},\ and\
  \citenamefont {McGuire}}]{May2016}%
  \BibitemOpen
  \bibfield  {author} {\bibinfo {author} {\bibfnamefont {A.~F.}\ \bibnamefont
  {May}}, \bibinfo {author} {\bibfnamefont {S.}~\bibnamefont {Calder}},
  \bibinfo {author} {\bibfnamefont {C.}~\bibnamefont {Cantoni}}, \bibinfo
  {author} {\bibfnamefont {H.}~\bibnamefont {Cao}},\ and\ \bibinfo {author}
  {\bibfnamefont {M.~A.}\ \bibnamefont {McGuire}},\ }\bibfield  {title}
  {\bibinfo {title} {Magnetic structure and phase stability of the van der
  waals bonded ferromagnet $\mathrm{Fe_{3-x}GeTe_2}$},\ }\href
  {https://doi.org/10.1103/PhysRevB.93.014411} {\bibfield  {journal} {\bibinfo
  {journal} {Phys. Rev. B}\ }\textbf {\bibinfo {volume} {93}},\ \bibinfo
  {pages} {014411} (\bibinfo {year} {2016})}\BibitemShut {NoStop}%
\bibitem [{\citenamefont {Verchenko}\ \emph {et~al.}(2015)\citenamefont
  {Verchenko}, \citenamefont {Tsirlin}, \citenamefont {Sobolev}, \citenamefont
  {Presniakov},\ and\ \citenamefont {Shevelkov}}]{Verchenko2015}%
  \BibitemOpen
  \bibfield  {author} {\bibinfo {author} {\bibfnamefont {V.~Y.}\ \bibnamefont
  {Verchenko}}, \bibinfo {author} {\bibfnamefont {A.~A.}\ \bibnamefont
  {Tsirlin}}, \bibinfo {author} {\bibfnamefont {A.~V.}\ \bibnamefont
  {Sobolev}}, \bibinfo {author} {\bibfnamefont {I.~A.}\ \bibnamefont
  {Presniakov}},\ and\ \bibinfo {author} {\bibfnamefont {A.~V.}\ \bibnamefont
  {Shevelkov}},\ }\bibfield  {title} {\bibinfo {title} {Ferromagnetic order,
  strong magnetocrystalline anisotropy, and magnetocaloric effect in the
  layered telluride $\mathrm{Fe_{(3-\delta)}GeTe_2}$},\ }\href
  {https://doi.org/doi: 10.1021/acs.inorgchem.5b01260.} {\bibfield  {journal}
  {\bibinfo  {journal} {Inorg. Chem.}\ }\textbf {\bibinfo {volume} {54}},\
  \bibinfo {pages} {8598} (\bibinfo {year} {2015})}\BibitemShut {NoStop}%
\bibitem [{\citenamefont {Zhu}\ \emph {et~al.}(2016)\citenamefont {Zhu},
  \citenamefont {Janoschek}, \citenamefont {Chaves}, \citenamefont {Cezar},
  \citenamefont {Durakiewicz}, \citenamefont {Ronning}, \citenamefont {Sassa},
  \citenamefont {Mansson}, \citenamefont {Scott}, \citenamefont {Wakeham},
  \citenamefont {Bauer},\ and\ \citenamefont {Thompson}}]{Zhu2016}%
  \BibitemOpen
  \bibfield  {author} {\bibinfo {author} {\bibfnamefont {J.-X.}\ \bibnamefont
  {Zhu}}, \bibinfo {author} {\bibfnamefont {M.}~\bibnamefont {Janoschek}},
  \bibinfo {author} {\bibfnamefont {D.~S.}\ \bibnamefont {Chaves}}, \bibinfo
  {author} {\bibfnamefont {J.~C.}\ \bibnamefont {Cezar}}, \bibinfo {author}
  {\bibfnamefont {T.}~\bibnamefont {Durakiewicz}}, \bibinfo {author}
  {\bibfnamefont {F.}~\bibnamefont {Ronning}}, \bibinfo {author} {\bibfnamefont
  {Y.}~\bibnamefont {Sassa}}, \bibinfo {author} {\bibfnamefont
  {M.}~\bibnamefont {Mansson}}, \bibinfo {author} {\bibfnamefont {B.~L.}\
  \bibnamefont {Scott}}, \bibinfo {author} {\bibfnamefont {N.}~\bibnamefont
  {Wakeham}}, \bibinfo {author} {\bibfnamefont {E.~D.}\ \bibnamefont {Bauer}},\
  and\ \bibinfo {author} {\bibfnamefont {J.~D.}\ \bibnamefont {Thompson}},\
  }\bibfield  {title} {\bibinfo {title} {Electronic correlation and magnetism
  in the ferromagnetic metal $\mathrm{Fe_{3}GeTe_2}$},\ }\href
  {https://doi.org/10.1103/PhysRevB.93.144404} {\bibfield  {journal} {\bibinfo
  {journal} {Phys. Rev. B}\ }\textbf {\bibinfo {volume} {93}},\ \bibinfo
  {pages} {144404} (\bibinfo {year} {2016})}\BibitemShut {NoStop}%
\bibitem [{\citenamefont {León-Brito}\ \emph {et~al.}(2016)\citenamefont
  {León-Brito}, \citenamefont {Bauer}, \citenamefont {Ronning}, \citenamefont
  {Thompson},\ and\ \citenamefont {Movshovich}}]{Brito2016}%
  \BibitemOpen
  \bibfield  {author} {\bibinfo {author} {\bibfnamefont {N.}~\bibnamefont
  {León-Brito}}, \bibinfo {author} {\bibfnamefont {E.~D.}\ \bibnamefont
  {Bauer}}, \bibinfo {author} {\bibfnamefont {F.}~\bibnamefont {Ronning}},
  \bibinfo {author} {\bibfnamefont {J.~D.}\ \bibnamefont {Thompson}},\ and\
  \bibinfo {author} {\bibfnamefont {R.}~\bibnamefont {Movshovich}},\ }\bibfield
   {title} {\bibinfo {title} {Magnetic microstructure and magnetic properties
  of uniaxial itinerant ferromagnet $\mathrm{Fe_{3}GeTe_2}$},\ }\href
  {https://doi.org/10.1063/1.4961592} {\bibfield  {journal} {\bibinfo
  {journal} {J. Appl. Phys.}\ }\textbf {\bibinfo {volume} {120}},\ \bibinfo
  {pages} {083903} (\bibinfo {year} {2016})},\ \Eprint
  {https://arxiv.org/abs/https://doi.org/10.1063/1.4961592}
  {https://doi.org/10.1063/1.4961592} \BibitemShut {NoStop}%
\bibitem [{\citenamefont {Chikazumi}(1997)}]{Chikazumi1997}%
  \BibitemOpen
  \bibfield  {author} {\bibinfo {author} {\bibfnamefont {S.}~\bibnamefont
  {Chikazumi}},\ }\href@noop {} {\emph {\bibinfo {title} {Physics of
  Ferromagnetism}}}\ (\bibinfo  {publisher} {Oxford University Press},\
  \bibinfo {year} {1997})\BibitemShut {NoStop}%
\bibitem [{\citenamefont {Rodríguez-Carvajal}(1993)}]{Fullprof}%
  \BibitemOpen
  \bibfield  {author} {\bibinfo {author} {\bibfnamefont {J.}~\bibnamefont
  {Rodríguez-Carvajal}},\ }\bibfield  {title} {\bibinfo {title} {Recent
  advances in magnetic structure determination by neutron powder diffraction},\
  }\href {https://doi.org/https://doi.org/10.1016/0921-4526(93)90108-I}
  {\bibfield  {journal} {\bibinfo  {journal} {Physica B: Condens. Matter}\
  }\textbf {\bibinfo {volume} {192}},\ \bibinfo {pages} {55} (\bibinfo {year}
  {1993})}\BibitemShut {NoStop}%
\bibitem [{\citenamefont {Reisser}\ \emph {et~al.}(1995)\citenamefont
  {Reisser}, \citenamefont {Kremer},\ and\ \citenamefont {Simon}}]{Reisser1}%
  \BibitemOpen
  \bibfield  {author} {\bibinfo {author} {\bibfnamefont {R.}~\bibnamefont
  {Reisser}}, \bibinfo {author} {\bibfnamefont {R.~K.}\ \bibnamefont
  {Kremer}},\ and\ \bibinfo {author} {\bibfnamefont {A.}~\bibnamefont
  {Simon}},\ }\bibfield  {title} {\bibinfo {title} {Maagnetic phase-transition
  in the metal-rich rare-earth carbide halides $\mathrm{Gd_2XC}$ $\mathrm{(X =
  BR,I)}$},\ }\href {https://doi.org/10.1103/PhysRevB.52.3546} {\bibfield
  {journal} {\bibinfo  {journal} {Physical Review B}\ }\textbf {\bibinfo
  {volume} {52}},\ \bibinfo {pages} {3546} (\bibinfo {year}
  {1995})}\BibitemShut {NoStop}%
\bibitem [{\citenamefont {Reiser}\ \emph {et~al.}(1995)\citenamefont {Reiser},
  \citenamefont {Kremer},\ and\ \citenamefont {Simon}}]{Reisser2}%
  \BibitemOpen
  \bibfield  {author} {\bibinfo {author} {\bibfnamefont {R.}~\bibnamefont
  {Reiser}}, \bibinfo {author} {\bibfnamefont {R.~K.}\ \bibnamefont {Kremer}},\
  and\ \bibinfo {author} {\bibfnamefont {A.}~\bibnamefont {Simon}},\ }\bibfield
   {title} {\bibinfo {title} {3d-xy critical-behavior of the layered metal-rich
  halides gd2ife2, $\mathrm{Gd_2ICo_2}$ and $\mathrm{Gd_2BrFe_2}$},\ }\href
  {https://doi.org/10.1016/0921-4526(94)00273-X} {\bibfield  {journal}
  {\bibinfo  {journal} {Physica B}\ }\textbf {\bibinfo {volume} {204}},\
  \bibinfo {pages} {265} (\bibinfo {year} {1995})}\BibitemShut {NoStop}%
\bibitem [{\citenamefont {Liu}\ \emph {et~al.}(2017{\natexlab{a}})\citenamefont
  {Liu}, \citenamefont {Ivanovski},\ and\ \citenamefont {Petrovic}}]{Liu1}%
  \BibitemOpen
  \bibfield  {author} {\bibinfo {author} {\bibfnamefont {Y.}~\bibnamefont
  {Liu}}, \bibinfo {author} {\bibfnamefont {V.~N.}\ \bibnamefont {Ivanovski}},\
  and\ \bibinfo {author} {\bibfnamefont {C.}~\bibnamefont {Petrovic}},\
  }\bibfield  {title} {\bibinfo {title} {Critical behavior of the van derwaals
  bonded ferromagnet $\mathrm{Fe_{2.72}GeTe_2}$},\ }\bibfield  {journal}
  {\bibinfo  {journal} {Physical Review B}\ }\textbf {\bibinfo {volume} {96}},\
  \href {https://doi.org/10.1103/PhysRevB.96.144429}
  {10.1103/PhysRevB.96.144429} (\bibinfo {year}
  {2017}{\natexlab{a}})\BibitemShut {NoStop}%
\bibitem [{\citenamefont {Liu}\ \emph {et~al.}(2017{\natexlab{b}})\citenamefont
  {Liu}, \citenamefont {Zou}, \citenamefont {Zhou}, \citenamefont {Zhang},
  \citenamefont {Wang}, \citenamefont {Li}, \citenamefont {Qu},\ and\
  \citenamefont {Zhang}}]{Liu2}%
  \BibitemOpen
  \bibfield  {author} {\bibinfo {author} {\bibfnamefont {B.}~\bibnamefont
  {Liu}}, \bibinfo {author} {\bibfnamefont {Y.}~\bibnamefont {Zou}}, \bibinfo
  {author} {\bibfnamefont {S.}~\bibnamefont {Zhou}}, \bibinfo {author}
  {\bibfnamefont {L.}~\bibnamefont {Zhang}}, \bibinfo {author} {\bibfnamefont
  {Z.}~\bibnamefont {Wang}}, \bibinfo {author} {\bibfnamefont {H.}~\bibnamefont
  {Li}}, \bibinfo {author} {\bibfnamefont {Z.}~\bibnamefont {Qu}},\ and\
  \bibinfo {author} {\bibfnamefont {Y.}~\bibnamefont {Zhang}},\ }\bibfield
  {title} {\bibinfo {title} {Critical behavior of the van der waals bonded high
  $\mathrm{T_C}$ ferromagnet $\mathrm{Fe_{2.72}GeTe_2}$},\ }\bibfield
  {journal} {\bibinfo  {journal} {Scientific Reports}\ }\textbf {\bibinfo
  {volume} {7}},\ \href {https://doi.org/10.1038/s41598-017-06671-5}
  {10.1038/s41598-017-06671-5} (\bibinfo {year}
  {2017}{\natexlab{b}})\BibitemShut {NoStop}%
\bibitem [{\citenamefont {Touloukian}\ \emph {et~al.}(1977)\citenamefont
  {Touloukian}, \citenamefont {Kirby}, \citenamefont {Taylor},\ and\
  \citenamefont {Lee}}]{Touloukian}%
  \BibitemOpen
  \bibfield  {author} {\bibinfo {author} {\bibfnamefont {Y.~S.}\ \bibnamefont
  {Touloukian}}, \bibinfo {author} {\bibfnamefont {R.~K.}\ \bibnamefont
  {Kirby}}, \bibinfo {author} {\bibfnamefont {E.~R.}\ \bibnamefont {Taylor}},\
  and\ \bibinfo {author} {\bibfnamefont {T.~Y.~R.}\ \bibnamefont {Lee}},\
  }\href@noop {} {\emph {\bibinfo {title} {Thermophysical Properties of Matter
  - the TPRC Data Series. Volume 13. Thermal Expansion - Nonmetallic Solids}}}\
  (\bibinfo  {publisher} {John Wiley and Sons Ltd},\ \bibinfo {year}
  {1977})\BibitemShut {NoStop}%
\bibitem [{\citenamefont {Tolman}(1938)}]{Tolman1938}%
  \BibitemOpen
  \bibfield  {author} {\bibinfo {author} {\bibfnamefont {R.~C.}\ \bibnamefont
  {Tolman}},\ }\href@noop {} {\emph {\bibinfo {title} {The Principles of
  Statistical Mechanics}}}\ (\bibinfo  {publisher} {Oxford University Press,
  Oxford},\ \bibinfo {year} {1938})\BibitemShut {NoStop}%
\bibitem [{\citenamefont {Kim}\ \emph {et~al.}(2021)\citenamefont {Kim},
  \citenamefont {Lee}, \citenamefont {Jang}, \citenamefont {Kim},\ and\
  \citenamefont {Shim}}]{Kim2021}%
  \BibitemOpen
  \bibfield  {author} {\bibinfo {author} {\bibfnamefont {D.}~\bibnamefont
  {Kim}}, \bibinfo {author} {\bibfnamefont {C.}~\bibnamefont {Lee}}, \bibinfo
  {author} {\bibfnamefont {B.~G.}\ \bibnamefont {Jang}}, \bibinfo {author}
  {\bibfnamefont {K.}~\bibnamefont {Kim}},\ and\ \bibinfo {author}
  {\bibfnamefont {J.~H.}\ \bibnamefont {Shim}},\ }\bibfield  {title} {\bibinfo
  {title} {Drastic change of magnetic anisotropy in $\mathrm{Fe_{3}GeTe_2}$ and
  $\mathrm{Fe_{4}GeTe_2}$ monolayers under electric field studied by density
  functional theory},\ }\href {https://doi.org/10.1038/s41598-021-96639-3}
  {\bibfield  {journal} {\bibinfo  {journal} {Scientific Reports}\ }\textbf
  {\bibinfo {volume} {11}},\ \bibinfo {pages} {11567} (\bibinfo {year}
  {2021})},\ \bibinfo {note} {pMID: 35442017},\ \Eprint
  {https://arxiv.org/abs/https://doi.org/10.1021/acsnano.1c09150}
  {https://doi.org/10.1021/acsnano.1c09150} \BibitemShut {NoStop}%
\bibitem [{\citenamefont {Dang}\ \emph {et~al.}(2023)\citenamefont {Dang},
  \citenamefont {Kozlenko}, \citenamefont {Lis}, \citenamefont {Kichanov},
  \citenamefont {Lukin}, \citenamefont {Golosova}, \citenamefont {Savenko},
  \citenamefont {Duong}, \citenamefont {Phan}, \citenamefont {Tran},\ and\
  \citenamefont {Phan}}]{Dang2023}%
  \BibitemOpen
  \bibfield  {author} {\bibinfo {author} {\bibfnamefont {N.-T.}\ \bibnamefont
  {Dang}}, \bibinfo {author} {\bibfnamefont {D.~P.}\ \bibnamefont {Kozlenko}},
  \bibinfo {author} {\bibfnamefont {O.~N.}\ \bibnamefont {Lis}}, \bibinfo
  {author} {\bibfnamefont {S.~E.}\ \bibnamefont {Kichanov}}, \bibinfo {author}
  {\bibfnamefont {Y.~V.}\ \bibnamefont {Lukin}}, \bibinfo {author}
  {\bibfnamefont {N.~O.}\ \bibnamefont {Golosova}}, \bibinfo {author}
  {\bibfnamefont {B.~N.}\ \bibnamefont {Savenko}}, \bibinfo {author}
  {\bibfnamefont {D.-L.}\ \bibnamefont {Duong}}, \bibinfo {author}
  {\bibfnamefont {T.-L.}\ \bibnamefont {Phan}}, \bibinfo {author}
  {\bibfnamefont {T.-A.}\ \bibnamefont {Tran}},\ and\ \bibinfo {author}
  {\bibfnamefont {M.-H.}\ \bibnamefont {Phan}},\ }\bibfield  {title} {\bibinfo
  {title} {High pressure-driven magnetic disorder and structural transformation
  in $\mathrm{Fe_{3}GeTe_2}$: Emergence of a magnetic quantum critical point},\
  }\href {https://doi.org/https://doi.org/10.1002/advs.202206842} {\bibfield
  {journal} {\bibinfo  {journal} {Adv. Sci.}\ ,\ \bibinfo {pages} {2206842}}
  (\bibinfo {year} {2023})},\ \Eprint
  {https://arxiv.org/abs/https://onlinelibrary.wiley.com/doi/pdf/
  10.1002/advs.202206842} {https://onlinelibrary.wiley.com/doi/pdf/
  10.1002/advs.202206842} \BibitemShut {NoStop}%
\bibitem [{\citenamefont {O'Hara}\ \emph {et~al.}(2020)\citenamefont {O'Hara},
  \citenamefont {Brubaker}, \citenamefont {Stillwell}, \citenamefont
  {O'Bannon}, \citenamefont {Baker}, \citenamefont {Weber}, \citenamefont
  {Aji}, \citenamefont {Goldberger}, \citenamefont {Kawakami}, \citenamefont
  {Zieve}, \citenamefont {Jeffries},\ and\ \citenamefont {McCall}}]{Ohara2020}%
  \BibitemOpen
  \bibfield  {author} {\bibinfo {author} {\bibfnamefont {D.~J.}\ \bibnamefont
  {O'Hara}}, \bibinfo {author} {\bibfnamefont {Z.~E.}\ \bibnamefont
  {Brubaker}}, \bibinfo {author} {\bibfnamefont {R.~L.}\ \bibnamefont
  {Stillwell}}, \bibinfo {author} {\bibfnamefont {E.~F.}\ \bibnamefont
  {O'Bannon}}, \bibinfo {author} {\bibfnamefont {A.~A.}\ \bibnamefont {Baker}},
  \bibinfo {author} {\bibfnamefont {D.}~\bibnamefont {Weber}}, \bibinfo
  {author} {\bibfnamefont {L.~B.~B.}\ \bibnamefont {Aji}}, \bibinfo {author}
  {\bibfnamefont {J.~E.}\ \bibnamefont {Goldberger}}, \bibinfo {author}
  {\bibfnamefont {R.~K.}\ \bibnamefont {Kawakami}}, \bibinfo {author}
  {\bibfnamefont {R.~J.}\ \bibnamefont {Zieve}}, \bibinfo {author}
  {\bibfnamefont {J.~R.}\ \bibnamefont {Jeffries}},\ and\ \bibinfo {author}
  {\bibfnamefont {S.~K.}\ \bibnamefont {McCall}},\ }\bibfield  {title}
  {\bibinfo {title} {Suppression of magnetic ordering in fe-deficient
  $\mathrm{Fe_{3-x}GeTe_2}$ from application of pressure},\ }\href
  {https://doi.org/10.1103/PhysRevB.102.054405} {\bibfield  {journal} {\bibinfo
   {journal} {Phys. Rev. B}\ }\textbf {\bibinfo {volume} {102}},\ \bibinfo
  {pages} {054405} (\bibinfo {year} {2020})}\BibitemShut {NoStop}%
\bibitem [{\citenamefont {De~Jongh}\ and\ \citenamefont
  {Miedema}(2001)}]{Miedema2001}%
  \BibitemOpen
  \bibfield  {author} {\bibinfo {author} {\bibfnamefont {L.~J.}\ \bibnamefont
  {De~Jongh}}\ and\ \bibinfo {author} {\bibfnamefont {A.~R.}\ \bibnamefont
  {Miedema}},\ }\bibfield  {title} {\bibinfo {title} {Experiments on simple
  magnetic model systems},\ }\href {https://doi.org/10.1080/00018730110101412}
  {\bibfield  {journal} {\bibinfo  {journal} {Adv. Phys.}\ }\textbf {\bibinfo
  {volume} {50}},\ \bibinfo {pages} {947} (\bibinfo {year} {2001})},\ \Eprint
  {https://arxiv.org/abs/https://doi.org/10.1080/00018730110101412}
  {https://doi.org/10.1080/00018730110101412} \BibitemShut {NoStop}%
\bibitem [{\citenamefont {Bao}\ \emph {et~al.}(2022)\citenamefont {Bao},
  \citenamefont {Wang}, \citenamefont {Shangguan}, \citenamefont {Cai},
  \citenamefont {Dong}, \citenamefont {Huang}, \citenamefont {Si},
  \citenamefont {Ma}, \citenamefont {Kajimoto}, \citenamefont {Ikeuchi},
  \citenamefont {Yano}, \citenamefont {Yu}, \citenamefont {Wan}, \citenamefont
  {Li},\ and\ \citenamefont {Wen}}]{SongBao2022}%
  \BibitemOpen
  \bibfield  {author} {\bibinfo {author} {\bibfnamefont {S.}~\bibnamefont
  {Bao}}, \bibinfo {author} {\bibfnamefont {W.}~\bibnamefont {Wang}}, \bibinfo
  {author} {\bibfnamefont {Y.}~\bibnamefont {Shangguan}}, \bibinfo {author}
  {\bibfnamefont {Z.}~\bibnamefont {Cai}}, \bibinfo {author} {\bibfnamefont
  {Z.-Y.}\ \bibnamefont {Dong}}, \bibinfo {author} {\bibfnamefont
  {Z.}~\bibnamefont {Huang}}, \bibinfo {author} {\bibfnamefont
  {W.}~\bibnamefont {Si}}, \bibinfo {author} {\bibfnamefont {Z.}~\bibnamefont
  {Ma}}, \bibinfo {author} {\bibfnamefont {R.}~\bibnamefont {Kajimoto}},
  \bibinfo {author} {\bibfnamefont {K.}~\bibnamefont {Ikeuchi}}, \bibinfo
  {author} {\bibfnamefont {S.-i.}\ \bibnamefont {Yano}}, \bibinfo {author}
  {\bibfnamefont {S.-L.}\ \bibnamefont {Yu}}, \bibinfo {author} {\bibfnamefont
  {X.}~\bibnamefont {Wan}}, \bibinfo {author} {\bibfnamefont {J.-X.}\
  \bibnamefont {Li}},\ and\ \bibinfo {author} {\bibfnamefont {J.}~\bibnamefont
  {Wen}},\ }\bibfield  {title} {\bibinfo {title} {Neutron spectroscopy evidence
  on the dual nature of magnetic excitations in a van der waals metallic
  ferromagnet $\mathrm{Fe_{2.72}GeTe_2}$},\ }\href
  {https://doi.org/10.1103/PhysRevX.12.011022} {\bibfield  {journal} {\bibinfo
  {journal} {Phys. Rev. X}\ }\textbf {\bibinfo {volume} {12}},\ \bibinfo
  {pages} {011022} (\bibinfo {year} {2022})}\BibitemShut {NoStop}%
\bibitem [{\citenamefont {Grosu}\ and\ \citenamefont
  {Crisan}(2020)}]{Grosu2020}%
  \BibitemOpen
  \bibfield  {author} {\bibinfo {author} {\bibfnamefont {I.}~\bibnamefont
  {Grosu}}\ and\ \bibinfo {author} {\bibfnamefont {M.}~\bibnamefont {Crisan}},\
  }\bibfield  {title} {\bibinfo {title} {Ferromagnetic order in two-dimensional
  spin systems with dipolar interaction},\ }\href
  {https://doi.org/10.1007/s10948-019-05320-4} {\bibfield  {journal} {\bibinfo
  {journal} {J. Supercond. Nov. Magn.}\ }\textbf {\bibinfo {volume} {33}},\
  \bibinfo {pages} {1073} (\bibinfo {year} {2020})}\BibitemShut {NoStop}%
\end{thebibliography}

\end{document}